\documentclass[a4paper,11pt]{article}

\usepackage{jcappub} 

\usepackage{graphics}

\usepackage{amssymb}

\usepackage{amsmath}

\title{\boldmath First Results on Dark Matter Annihilation in the Sun using the ANTARES Neutrino Telescope}

\author[a]{S.~Adri\'an-Mart\'inez}
\author[b]{I. Al Samarai}
\author[c]{A. Albert}
\author[d]{M.~Andr\'e}
\author[e]{M. Anghinolfi}
\author[f]{G. Anton}
\author[f]{L. Anton}
\author[g]{S. Anvar}
\author[a]{M. Ardid}
\author[h,1]{T.~Astraatmadja\note{Also at University of Leiden, the Netherlands}}
% ...

\affiliation[a]{\scriptsize{Institut d'Investigaci\'o per a la Gesti\'o Integrada de les Zones Costaneres (IGIC) - Universitat Polit\`ecnica de Val\`encia. C/  Paranimf 1 , 46730 Gandia, Spain.}}
\affiliation[b]{\scriptsize{CPPM, Aix-Marseille Universit\'e, CNRS/IN2P3, Marseille, France}}
\affiliation[c]{\scriptsize{GRPHE - Institut universitaire de technologie de Colmar, 34 rue du Grillenbreit BP 50568 - 68008 Colmar, France }}
\affiliation[d]{\scriptsize{Technical University of Catalonia, Laboratory of Applied Bioacoustics, Rambla Exposici\'o,08800 Vilanova i la Geltr\'u,Barcelona, Spain}}
\affiliation[e]{\scriptsize{INFN - Sezione di Genova, Via Dodecaneso 33, 16146 Genova, Italy}}
\affiliation[f]{\scriptsize{Friedrich-Alexander-Universit\"at Erlangen-N\"urnberg, Erlangen Centre for Astroparticle Physics, Erwin-Rommel-Str. 1, 91058 Erlangen, Germany}}
\affiliation[g]{\scriptsize{Direction des Sciences de la Mati\`ere - Institut de recherche sur les lois fondamentales de l'Univers - Service d'Electronique des D\'etecteurs et d'Informatique, CEA Saclay, 91191 Gif-sur-Yvette Cedex, France}}
\affiliation[h]{\scriptsize{Nikhef, Science Park,  Amsterdam, The Netherlands}}
% ...

\abstract{A search for high-energy neutrinos coming from the     
direction of the Sun has been performed using the data recorded by 
the ANTARES neutrino telescope during 2007 and 2008. The neutrino selection 
criteria have been chosen to maximize the selection of possible signals produced by   
the self-annihilation of weakly interacting massive particles	      
accumulated in the centre of the Sun with respect to the atmospheric background. After data unblinding, 
the number of neutrinos observed towards the Sun was found to be compatible with background 
expectations. The 90\% CL upper limits in terms of spin-dependent and spin-independent 
WIMP-proton cross-sections are derived and compared to predictions of two supersymmetric 
models, CMSSM and MSSM-7. The ANTARES limits are comparable with those obtained by other 
neutrino observatories and are more stringent than those obtained by direct search 
experiments for the spin-dependent WIMP-proton cross-section in the case of hard 
self-annihilation channels ({\rm $W^{+}W^{-}$, $\tau^{+}\tau^{-}$}).}


\begin{document}
\maketitle
\flushbottom

\section{Introduction}
\label{introduction}

There is compelling evidence from cosmology and astrophysics that about 
$83$~\% of the matter in the Universe is non-baryonic, non-relativistic and 
does not interact electromagnetically --- the so-called dark matter~\cite{darkmatter,pdg2012}. 
Much of this evidence comes from the internal dynamics of galaxy clusters~\cite{galclust}, 
the rotation curves of galaxies~\cite{galrot}, the observations from weak lensing 
($\rm 1E 0657-558$)~\cite{bulclust}, but also from the Cosmic Microwave Background (CMB), 
the large scale structure formation and type Ia supernovae. The determination of the 
relic density of cold dark matter (CDM) in the Universe is $\rm \Omega_{CDM}h^{2} = 0.1120 \pm 0.0056$ 
using observations of the CMB~\cite{wmap7yr}. A popular hypothesis is that dark 
matter is made of Weakly Interacting Massive Particles (WIMPs) that are embedded in the
visible baryonic part of galaxies and surround them in the form of a halo. There are 
a variety of candidates for WIMPs, among which those provided by theories based on 
supersymmetry (SUSY) attract a great deal of interest. In some classes of the minimal 
supersymmetric extension of the Standard Model (MSSM), the lightest supersymmetric particle 
(LSP) is stable thanks to the conservation of R-parity that forbids its decay to standard 
particles. Consequently, the LSP can only annihilate in pairs, making it a good WIMP candidate 
for dark matter~\cite{lsp0,lsp1}. In these models, high-energy neutrinos are produced from the 
decay of the LSPs' self-annihilation products. Two simplified versions of the MSSM model are 
considered in this paper, the constrained MSSM (CMSSM)~\cite{CMSSM} and the low-energy 
phenomenological model MSSM-7~\cite{MSSM7}. Both have a neutralino as the LSP.

The search for WIMPs can be performed either directly by recording the recoil energy of 
nuclei when WIMPs scatter off them in suitable detectors, or indirectly. The indirect approach, 
which is adopted here, exploits a radiation signature (gamma-ray, synchroton, positron, anti-proton 
or neutrino flux) produced by the self-annihilations of WIMPs accumulated in astrophysical objects 
such as the galactic halo, the Sun or the Earth~\cite{indirectdm}.

For the case of the Sun, dealt with in the paper, WIMPs can scatter elastically and become gravitationally 
trapped in its core. Here, the self-annihilation rate reaches a maximum when in equilibrium with the capture 
rate over the age of the Solar System~\cite{dmsun}. The WIMPs self-annihilate to Standard Model (SM) 
particles whose decay or hadronisation give rise to the production of energetic neutrinos which can 
escape from the Sun and be detected by neutrino telescopes on the Earth. The accumulation of WIMPs in the 
Sun must have taken place during a large period of time and therefore a very wide region in the Galaxy must 
have contributed, thereby reducing the dependence of the overall capture on the detailed sub-structures of the 
dark matter halo distribution. Moreover, high-energy neutrinos (above several GeV) coming from the Sun 
could not be explained by other known astrophysical processes.

\bigskip
In this paper an indirect search for dark matter by looking for high-energy neutrinos coming from the Sun, 
using the 2007-2008 data recorded by the ANTARES neutrino telescope, is reported. The layout of the paper 
is as follows. In Section~\ref{antares}, the main features of the ANTARES neutrino telescope 
and the reconstruction algorithm used in this work are described. 
In Section~\ref{simulation}, the Monte Carlo simulation of the WIMP signal, the background 
expected from atmospheric muons and neutrinos, and the grid scan performed to explore the 
parameter space of the CMSSM and MSSM-7 models are reported. In Section~\ref{optimisation}, 
the method used to optimise the selection of the neutrino events is described. Finally, 
the results obtained are discussed in Section~\ref{results}, where limits on the 
neutrino flux are derived from the absence of a signal coming from the Sun's direction. The 
corresponding limits on the spin-dependent and the spin-independent WIMP-proton cross-sections 
are obtained and compared to the predictions of the CMSSM and MSSM-7 theoretical 
models.   


\section{The ANTARES Neutrino Telescope}
\label{antares}

ANTARES is the first undersea neutrino telescope and the largest of its kind in the Northern 
Hemisphere~\cite{antares}. It is located between 2475 m (seabed) and 2025 m below the Mediterranean Sea level, 
$40$ km offshore from Toulon (France) at $42^{\circ} 48$' N and $6^{\circ} 10$' E. The telescope 
consists of $12$ detection lines with 25 storeys each. A standard storey includes 
three optical modules (OMs)~\cite{OM} each housing a 10-inch photomultiplier~\cite{PMT} and a local control 
module that contains the electronics~\cite{frontend, DAQ}. The OMs are orientated 45$^{\circ}$ 
downwards in order to optimise their acceptance to upgoing light and to avoid the effect of 
sedimentation and biofouling~\cite{biofouling}. The length of a line is 450 m and the horizontal 
distance between neighbouring lines is 60-75 m. In one of the lines, the upper storeys are dedicated 
to a test system for acoustic neutrino detection~\cite{amadeus}. Similar acoustic devices are also 
installed in an additional line that contains instrumentation aimed to measure environmental 
parameters~\cite{instrumentation}. The location of the active components of the lines is known 
better than 10~cm by a combination of tiltmeters and compasses in each storey and a series of
acoustic transceivers (emitters and receivers) in certain storeys along the line and surrounding 
the telescope~\cite{alignment}. A common time reference is maintained in the full detector by means 
of a 25 MHz clock signal broadcast from shore. The time offsets of the individual optical modules 
are determined in dedicated calibration facilities onshore and regularly monitored in situ by means 
of optical beacons distributed at various points of the apparatus which emit short light pulses 
through the water~\cite{OBs}. This allows to reach a sub-nanosecond accuracy on the relative 
timing~\cite{timing}. Additional information on the detector can be found in 
Reference~\cite{antares}.

A high-energy muon (anti-)neutrino interacts in the matter below the detector producing a relativistic 
muon that can travel hundreds of metres and cross the detector or pass nearby. This muon induces 
Cherenkov light when travelling through the water, which is detected by the OMs. From the time and 
position information of the photons provided by the OMs, the direction of the muon is reconstructed 
and is well correlated to the neutrino direction. 

Data taking started with the first 5 lines of the detector installed in 2007. The full detector 
was completed in May 2008 and has been operating continuously ever since, except for some periods 
in which repair and maintenance operations have taken place. Other physics results using this data-taking 
period can be found elsewhere~\cite{diffuse,ps,monopole}.

\bigskip
A muon track is reconstructed from the position and time of the hits of the Cherenkov photons in the OMs. 
The reconstruction algorithm ~\cite{bbfit} is based on the minimisation of a $\rm \chi^{2}$-like quality parameter, 
$\rm Q$, which uses the differences between the expected and measured times of the detected photons plus a correction 
term that takes into account the effect of light absorption: 


\begin{equation}
\rm  Q = \sum^{Nhit}_{i=1} \left[ \frac{(t_{\gamma}-t_{i})^{2}}{\sigma^{2}_{i}} + 
\frac{A(a_{i})D(d_{\gamma})}{<a>d_{0}} \right] \, ,
\label{Qeq}
\end{equation}

\noindent where $\rm t_{\gamma}$ and $\rm t_{i}$ are respectively the expected and recorded arrival time of the 
photons from the track, and $\sigma^{2}_{i}$ is the timing variance. The second term takes into account the 
accumulation of high charges in storeys close to the track. This term uses the measured hit charge, $\rm a_{i}$, 
the average hit charge calculated from all hits which have been selected for the fit, $<a>$, and the calculated 
photon travel distance, $\rm d_{\gamma}$, together with a normalisation value, $\rm d_{0}$. The functions 
$\rm A(a_{i})$ and $\rm D(d_{\gamma})$ are discussed at length in Reference~\cite{bbfit}.

\begin{figure}[!t]
\begin{center}
\includegraphics[width=0.8\linewidth]{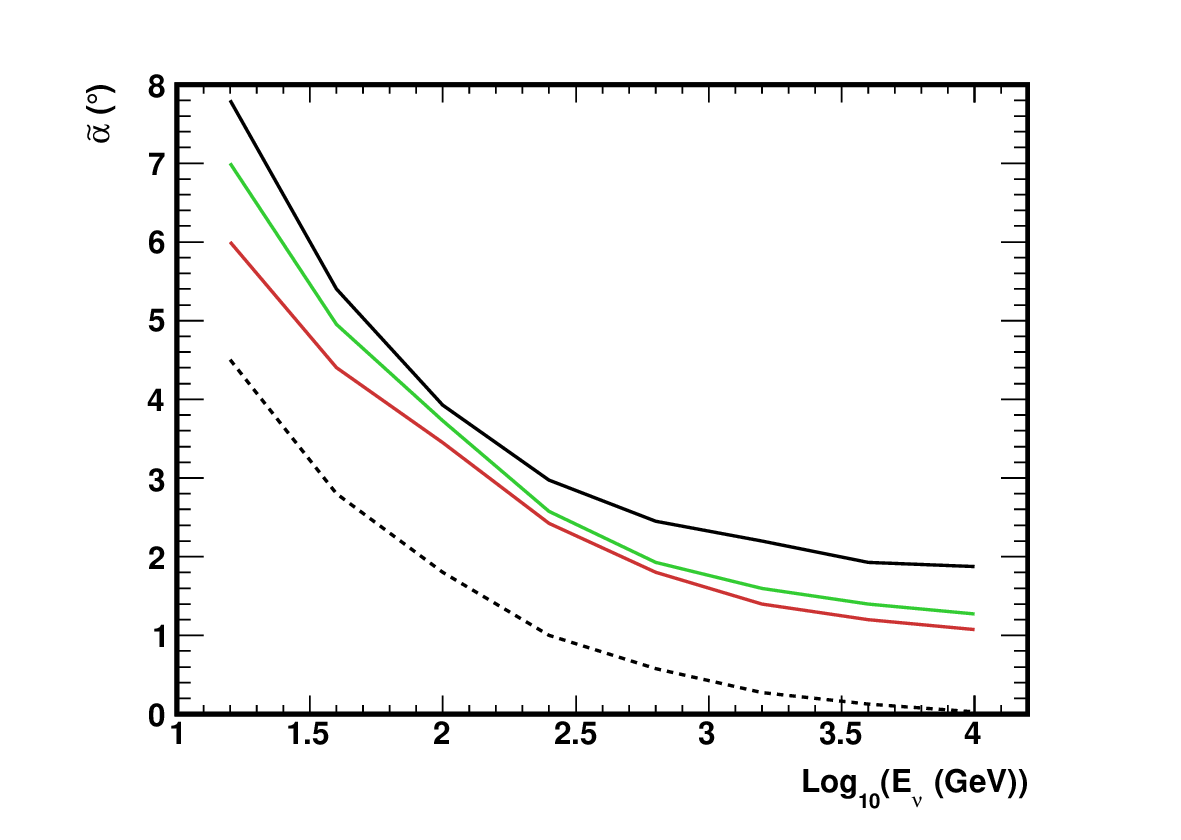}
\caption{Median angular error, $\tilde{\alpha}$ ($^{\circ}$), on the upgoing neutrino track (solid lines) in the energy 
range of interest $\rm E_{\nu} \leq 10$ TeV, for 5 (black), 9 (green), and 12 (red) line configuration of the detector 
(10 and 12 line angular resolution are identical). The black dashed line shows only the kinematic counterpart at the 
neutrino-muon vertex of interaction.}
\label{angresfig}
\end{center}
\end{figure}

Depending on the configuration of the detector (see Section ~\ref{optimisation}) and the muon (anti-) neutrino energy, 
this algorithm yields an angular resolution on the upgoing neutrino direction between $\rm 1$ and $\rm 7.8$ degrees as
illustrated by the Figure~\ref{angresfig}.


\begin{figure}[!t]
\begin{center}
\begin{minipage}[c]{.8\linewidth}
\includegraphics[width=\linewidth]{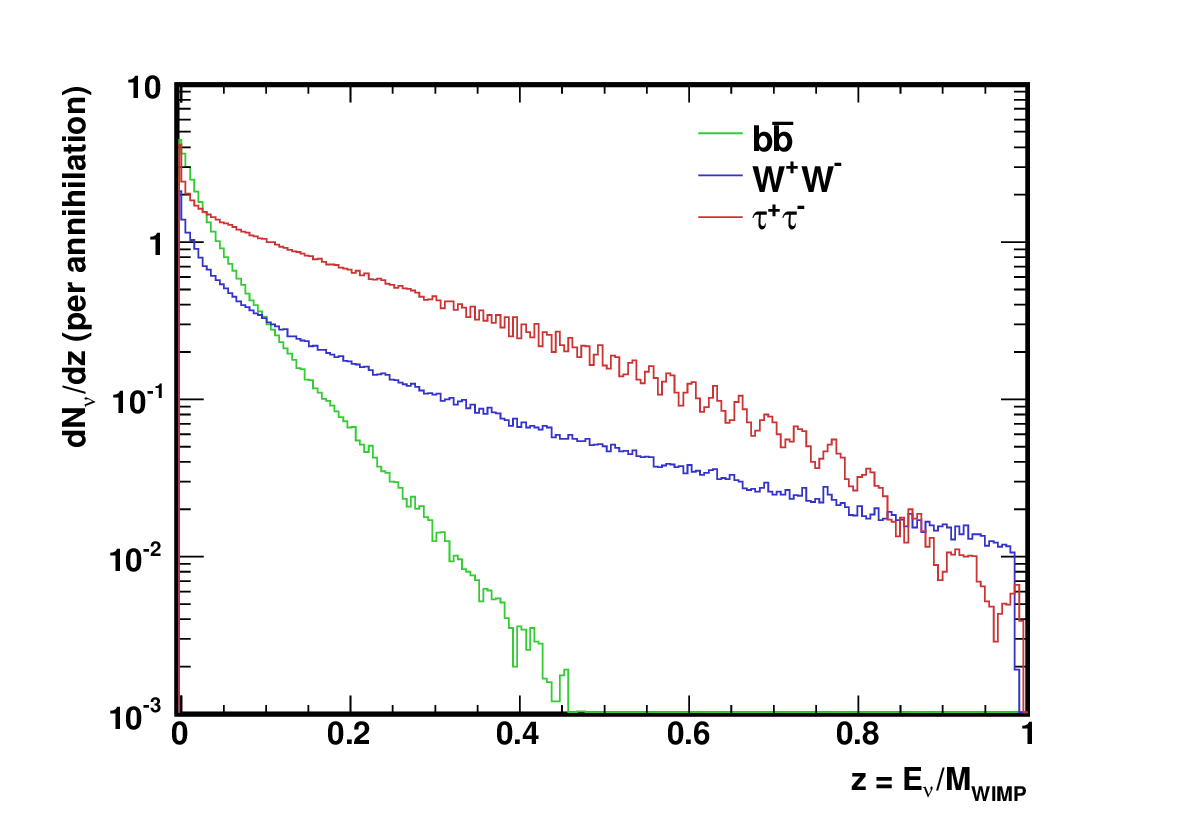} \\
\includegraphics[width=\linewidth]{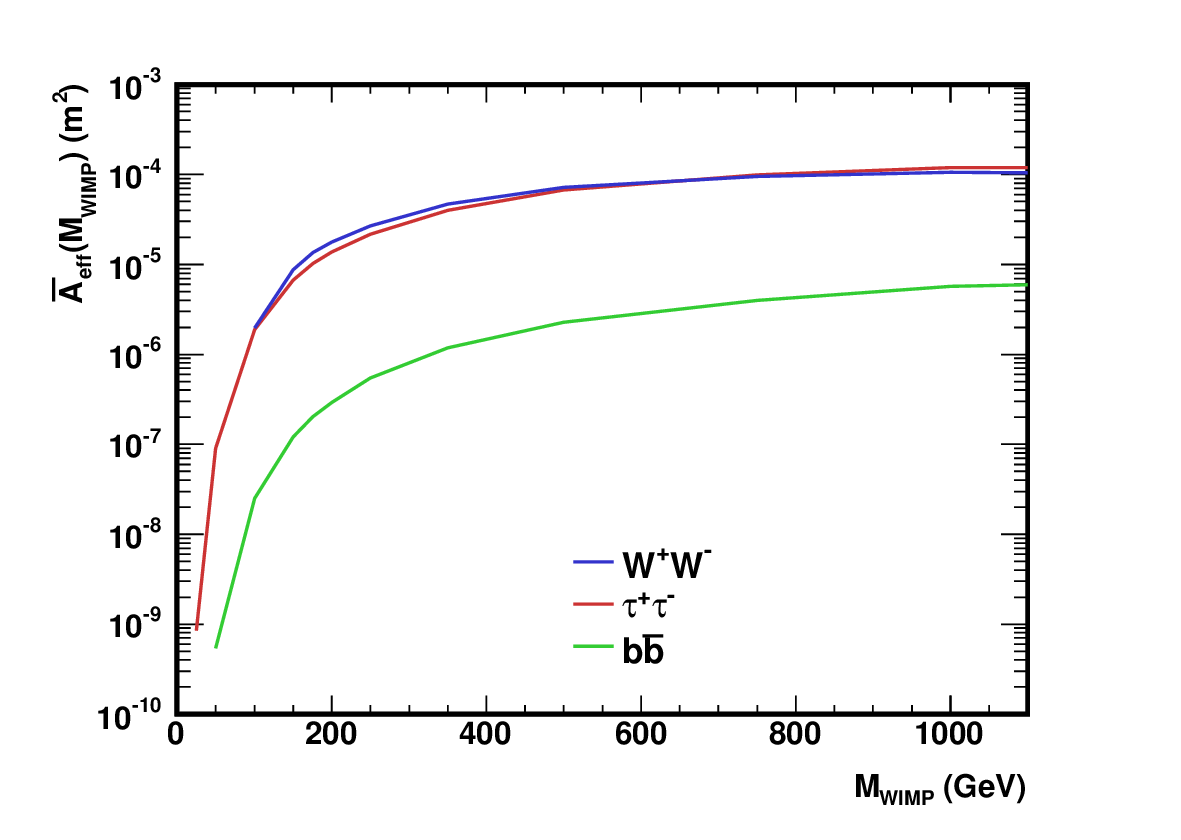} 
\end{minipage}
\caption{{\bf Top:} Distribution of the number of muon neutrinos at the surface of the Earth as a function of their 
energy normalised to the WIMP mass for the channels: $b\bar{b}$ (green), $W^{+}W^{-}$ (blue), $\tau^{+}\tau^{-}$ (red) for a WIMP mass 
$M_{\rm WIMP} = 350$ GeV, as an example. {\bf Bottom:} Examples of the averaged effective area $\rm \bar{A}_{eff}(M_{\rm WIMP})$ 
for the signal of WIMP self-annihilation inside the Sun, $b\bar{b}$ (green), $W^{+}W^{-}$ (blue) and $\tau^{+}\tau^{-}$ (red) 
channels. The detector is in a 12 line configuration and ($\rm Q_{cut}$,$\rm \Psi_{cut}$) = ($1.4$, $3^{\circ}$).}
\label{aeffmwimpfig}
\end{center}
\end{figure}

\section{Signal and background simulation}
\label{simulation}

The flux of neutrinos as a function of their energy arriving at the Earth's surface from the Sun's 
core is computed using the software package WimpSim~\cite{wimpsim} without theoretical assumptions 
concerning the dark matter model. The neutrinos resulting from the self-annihilation channels were 
simulated for $16$ different WIMP masses in the range from 50~GeV to 10~TeV.

\begin{figure}[!t]
\begin{center}
\includegraphics[width=0.8\linewidth]{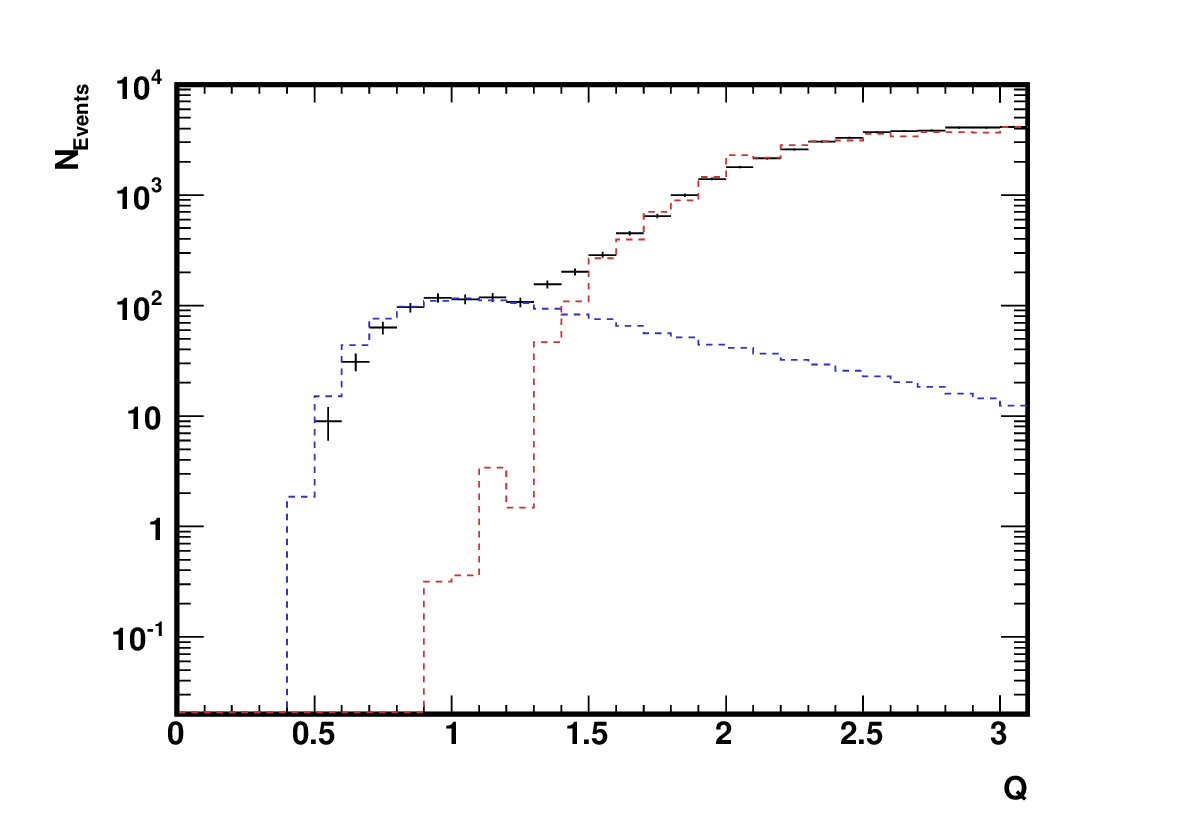}
\caption{Distribution of the track fit quality parameter, $\rm Q$. The blue and red dashed lines are, respectively, 
the expectations for atmospheric neutrino and muon events according to simulation and the black crosses are the 2007-2008 
data.}
\label{qcutfig}
\end{center}
\end{figure}

Three main self-annihilation channels are chosen as benchmarks for the lightest neutralino, $\rm \tilde{\chi}_{1}^{0}$, 
namely: a soft neutrino channel, $\rm \tilde{\chi}_{1}^{0}\tilde{\chi}_{1}^{0}\rightarrow b\bar{b}$, and two hard 
neutrino channels, $\rm \tilde{\chi}_{1}^{0}\tilde{\chi}_{1}^{0} \rightarrow W^{+}W^{-}$ and 
$\rm \tilde{\chi}_{1}^{0}\tilde{\chi}_{1}^{0} \rightarrow \tau^{+}\tau^{-}$. As the region in the SUSY parameter 
space determines which of these three channels is dominant, a 100\% branching ratio is assumed for all of them 
in order to explore the widest theoretical parameter space~\cite{abbasi, boliev, ambrosio, desai, ackermann}. 
The distribution of the number of muon neutrinos, $\rm dN_{\nu}/dz$, arriving at the Earth per pair of WIMPs 
self-annihilating in the Sun's core as a function of the energy ratio, $\rm z = E_{\nu}/M_{\rm WIMP}$, is shown in 
Figure~\ref{aeffmwimpfig} (top) for the channels $b\bar{b}$, $W^{+}W^{-}$ and $\tau^{+}\tau^{-}$ 
(equivalent spectra are determined for muon anti-neutrinos). In this simulation, oscillations among the 
three neutrino flavours (both in the Sun and during their flight to Earth) are taken into account, as well as 
$\nu$ absorption and $\tau$ lepton regeneration in the Sun's medium. 

\bigskip
The main backgrounds for cosmic neutrinos in a neutrino telescope are atmospheric muons and neutrinos, both 
produced in the interactions of cosmic rays with the Earth's atmosphere. Downgoing atmospheric muons dominate 
the trigger rate, which ranges from $3$ to $10$ Hz depending on the exact trigger conditions. They are simulated 
using Corsika~\cite{corsika}. Upgoing atmospheric neutrinos, which are recorded at a rate of $\sim$50 $\mu$Hz 
(about four per day), are simulated according to the parameterisation of the atmospheric $\rm \nu_{\mu}$ flux from 
Reference~\cite{bartol} in the energy range from 10~GeV to 10~PeV. The Cherenkov light produced in the vicinity 
of the detector is propagated taking into account light absorption and scattering in sea water~\cite{transmission}. 
The angular acceptance, quantum efficiency and other characteristics of the PMTs are taken from Reference~\cite{OM} 
and the overall geometry corresponds to the different layouts of the ANTARES detector during each data-taking period.

A source of background specific to this search is due to the interaction of cosmic rays with the Sun's corona. 
The interaction products may give rise to neutrinos in their decay. Using a simple parameterisation of the estimated 
$\rm \nu_\mu$ flux from Reference~\cite{hettlage} in the energy range from 10~GeV to 10~PeV, this background is found 
to amount to less than $0.4$~\% of the total atmospheric background in the direction of the Sun and therefore 
neglected.

\bigskip
To reduce the background from atmospheric muons, only upgoing events occurring during a period in which the Sun was 
below the horizon are kept. The residual contamination from misreconstructed downgoing muons is reduced using 
the quality parameter from Equation~\ref{Qeq}. Given the good agreement between data and simulated events as illustrated in 
Figure~\ref{qcutfig}, the simulated effective area is used to evaluate the expected signal (see Section~\ref{optimisation}). 
The expected background is estimated from the scrambled data (randomising the UTC time of the selected events) in order to 
minimise the effect of systematic uncertainties from the simulation.

\section{Optimisation of the event selection criteria}
\label{optimisation}

The data set used in this analysis comprises a total of $2693$ runs recorded between the $27^{\rm{th}}$ of 
January 2007 and the $31^{\rm{st}}$ of December 2008, corresponding to a total livetime of 294.6 days, without taking into 
account the period in which the Sun was below the horizon. The detector consisted of $5$ lines for most of $2007$ and of 
$9$, $10$ and $12$ lines during $2008$, with a corresponding total livetime of 134.6, 38.0, 39.0 and 83.0 days respectively. 

Only upgoing events are kept in the analysis. The track fit is required to use a number of hits greater than five 
in at least two lines in order to ensure a non-degenerate 5-parameter fit with an accurate reconstruction of 
the azimuth angle.

The UTC time of the events is uniformly randomised in the data-taking period in order to estimate the background 
in the Sun's direction from the data itself. The zenith and azimuth angles of the reconstructed tracks are kept so 
as to preserve the angular response of the detector in the optimisation of the selection criteria. This procedure 
provides a means to follow a data blinding strategy while using all the relevant information on the detector performance.

\bigskip
The values of the parameters used in the event selection criteria, the quality parameter, $\rm Q$ (see Equation~\ref{Qeq}), and the angular 
separation between tracks and the Sun's direction, $\rm \Psi$, are chosen so as to optimise the model rejection factor~\cite{mrf}. 
For each WIMP mass and each annihilation channel, the chosen individual values $\rm Q_{cut}$ and $\rm \Psi_{cut}$ are those that minimise 
the average 90\% confidence level (CL) upper limit on the $\rm \nu_{\mu}+\bar{\nu}_{\mu}$ flux, $\rm \overline{\Phi}_{\nu_{\mu}+\bar{\nu}_\mu}$, 
defined as

\begin{equation}
\rm{\overline{\Phi}_{\nu_{\mu}+\bar{\nu}_\mu} = \frac{\bar{\mu}^{90\%}}{\sum\limits_{i} \bar{A}_{eff}^{i}(M_{WIMP}) \times T_{eff}^{i}}} \, ,
\label{mrfeq}
\end{equation}

\noindent where the index $\rm i$ denotes the periods with different detector configurations 
(5, 9, 10 and 12 detection lines), $\rm \bar{\mu}^{90\%}$ is the average upper limit of the background at 
$90$\% CL computed using a Poisson distribution in the Feldman-Cousins approach~\cite{feldmancousins} (for consistency in the comparison with other 
neutrino experiments limits computation) and $\rm {T_{eff}^{i}}$ is the total livetime for each detector configuration. The effective area averaged 
over the neutrino energy, $\rm{\bar{A}_{eff}^{i}(M_{\rm WIMP})}$, is defined as:

\begin{equation}
\rm{\bar{A}_{eff}^{i}(M_{\rm WIMP}) =
\sum_{\nu,\bar{\nu}} \left ( \frac{\int_{E_{\nu}^{th}}^{M_{\rm WIMP}} A_{eff}^{i}(E_{\nu,\bar{\nu}}) \, \frac{dN_{\nu,\bar{\nu}}}{dE_{\nu,\bar{\nu}}} dE_{\nu,\bar{\nu}}}
{\int_{0}^{M_{\rm WIMP}}\frac{dN_{\nu}}{dE_{\nu}} dE_{\nu} \,+\, \frac{dN_{\bar{\nu}}}{dE_{\bar{\nu}}} dE_{\bar{\nu}}} \right )}  \, ,
\label{aeffeq}
\end{equation}

\begin{figure}[!t]
\begin{center}
\includegraphics[width=0.8\linewidth]{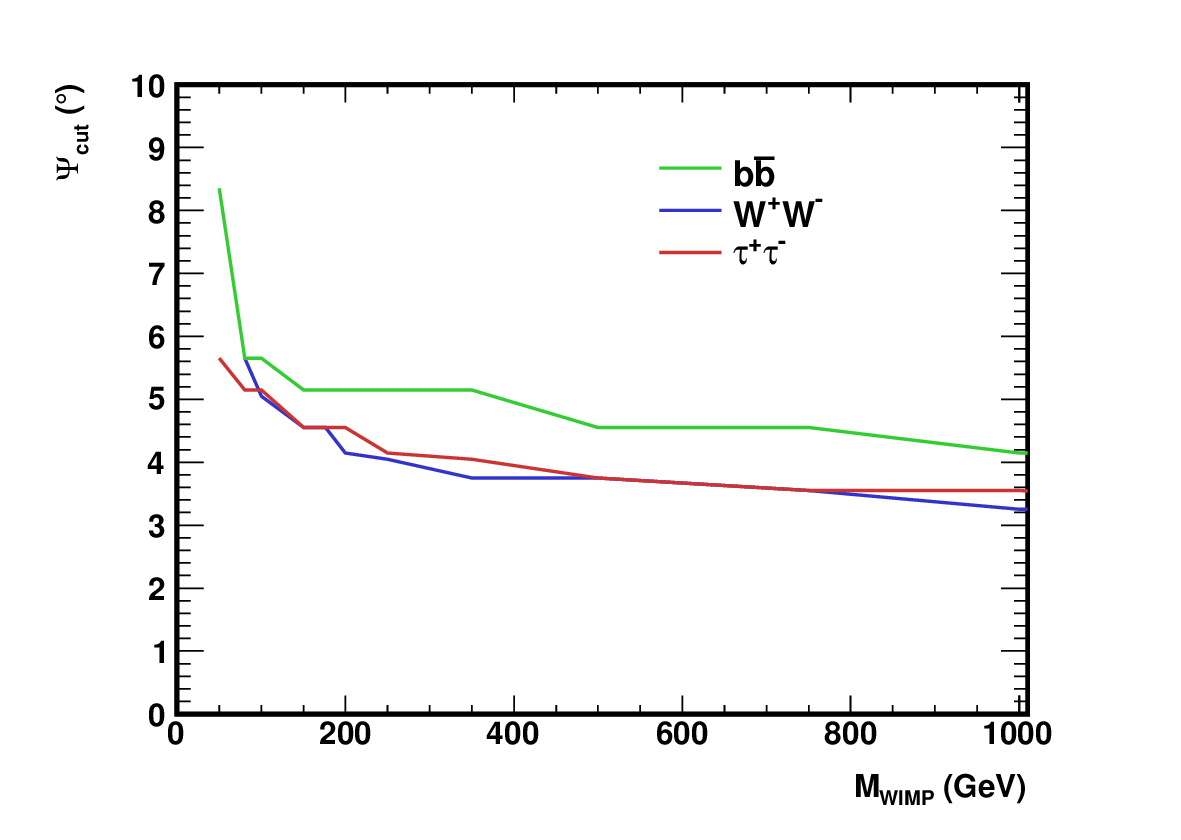} 
\end{center}
\caption{Optimum angular separation $\rm \Psi_{cut}$ between the muon tracks and the Sun's direction for $\rm Q_{cut} = 1.4$ 
as a function of the WIMP mass for the self-annihilation channels $b\bar{b}$ (green), $W^{+}W^{-}$ (blue) and 
$\tau^{+}\tau^{-}$ (red).}
\label{psifig}
\end{figure}

\begin{figure}[!t]
\begin{center}
\includegraphics[width=0.8\linewidth]{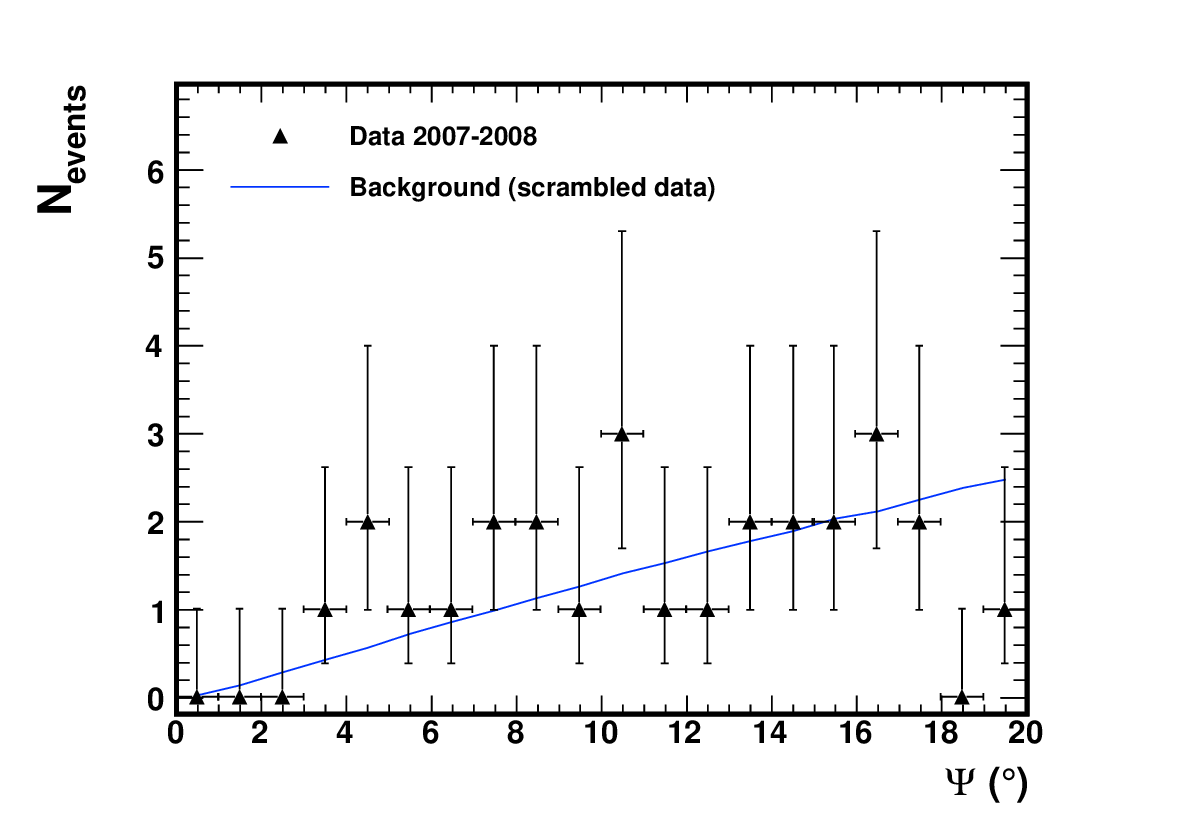} 
\end{center}
\caption{Differential distribution of the angular separation $\rm \Psi$ of the event tracks 
with respect to the Sun's direction for the expected background (solid blue line) compared to 
the data (black triangles). A $1\sigma$ Poisson uncertainty is shown for each data point.}
\label{databkgfig}
\end{figure}

\begin{figure}[!t]
\begin{center}
\begin{minipage}[c]{.8\linewidth}
\includegraphics[width=\linewidth]{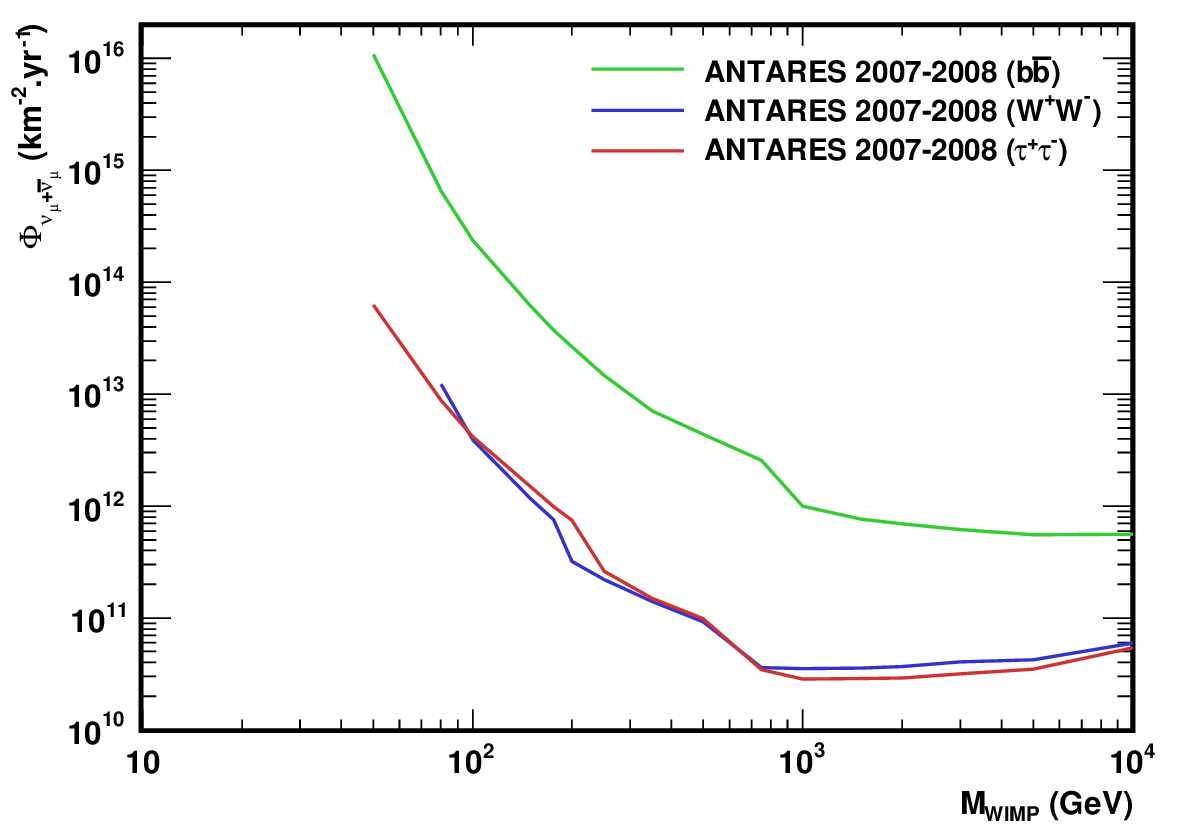} \\
\includegraphics[width=\linewidth]{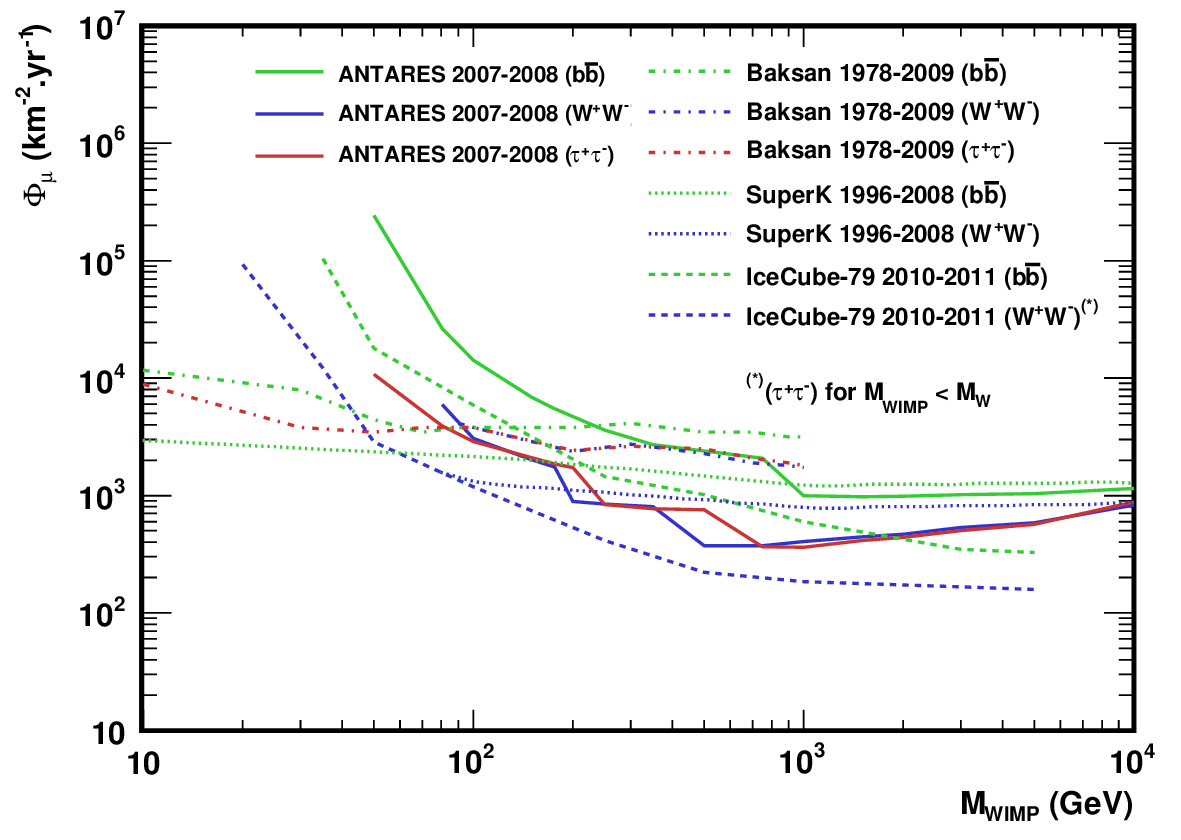}
\end{minipage}
\caption{{\bf Top:} $90$\% CL upper limits on the neutrino plus anti-neutrino flux 
as a function of the WIMP mass in the range $M_{\rm WIMP}\in$[$50$ GeV;$10$ TeV] for 
the three self-annihilation channels $b\bar{b}$ (green), $W^{+}W^{-}$ (blue), $\tau^{+}\tau^{-}$ (red). 
{\bf Bottom:} $90$\% CL upper limit on the muon flux as a function of the WIMP mass 
in the range $M_{\rm WIMP}\in$[$50$ GeV;$10$ TeV] for the three self-annihilation channels 
$b\bar{b}$ (green), $W^{+}W^{-}$ (blue) and $\tau^{+}\tau^{-}$ (red). The results 
from Baksan $1978-2009$~\cite{baksan} (dash-dotted lines), Super-Kamiokande $1996-2008$~\cite{superk} 
(dotted lines) and IceCube-$79$ $2010-2011$~\cite{icecube} (dashed lines) are also shown.}
\label{phinuphimulimitfig}
\end{center}
\end{figure}

\begin{figure}[h!]
\begin{center}
\begin{minipage}[c]{.78\linewidth}
\includegraphics[width=\linewidth]{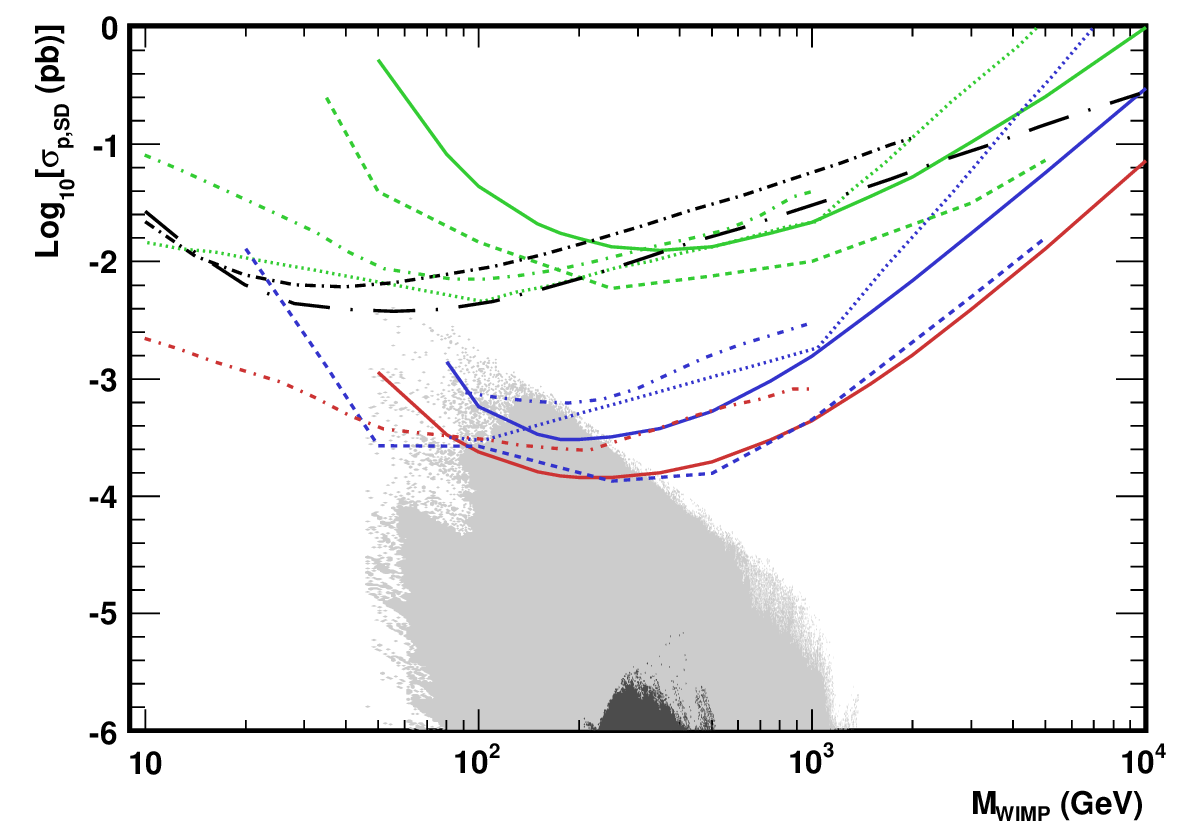} \\
\includegraphics[width=\linewidth]{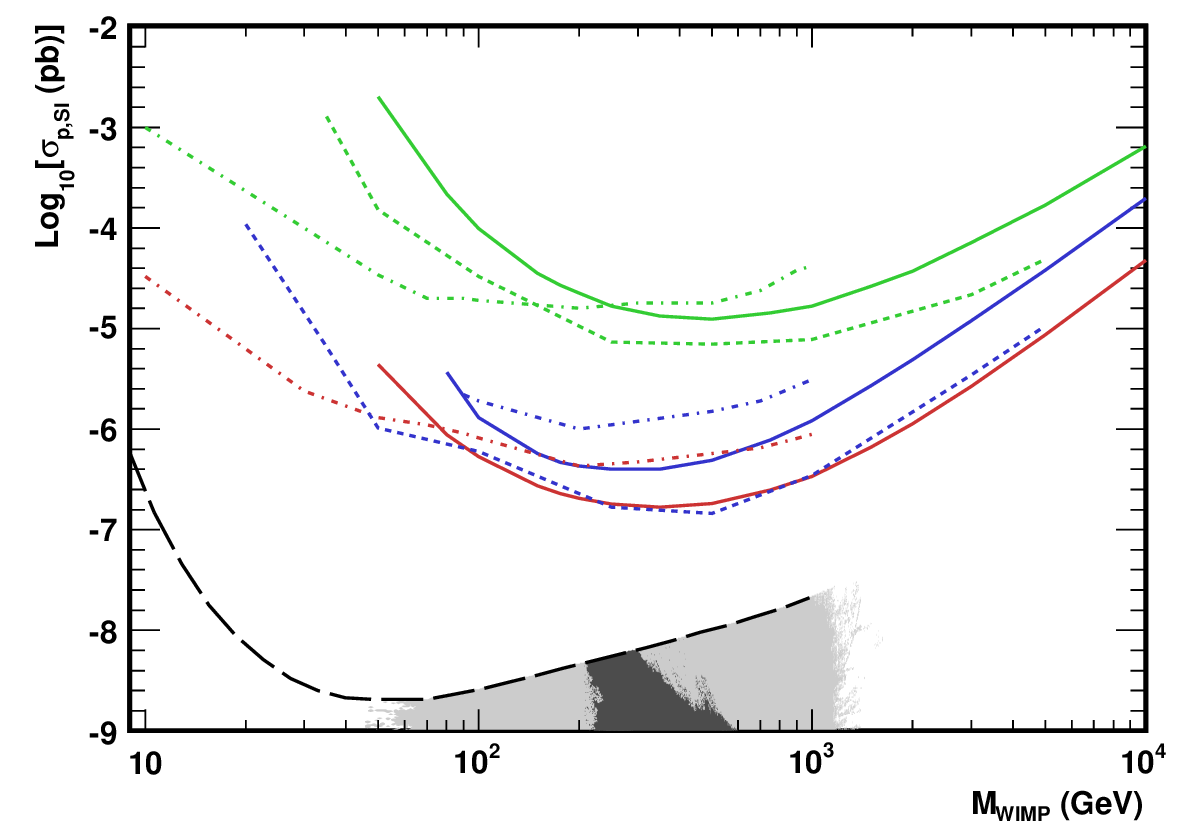}
\end{minipage}
\caption{$90$\% CL upper limits on the SD and SI WIMP-proton cross-sections (upper and lower plots, respectively) 
as a function of the WIMP mass, for the three self-annihilation channels: 
$b\bar{b}$ (green), $W^{+}W^{-}$ (blue) and $\tau^{+}\tau^{-}$ (red), for ANTARES 2007-2008 (solid line) compared 
to the results of other indirect search experiments: Baksan $1978-2009$~\cite{baksan} (dash-dotted lines), 
Super-Kamiokande $1996-2008$~\cite{superk} (dotted lines) and IceCube-$79$ $2010-2011$~\cite{icecube} (dashed lines) 
and the result of the most stringent direct search experiments (black): SIMPLE $2004-2011$~\cite{simple} 
(short dot-dashed line in upper plot), COUPP $2010-2011$~\cite{coupp} (long dot-dashed line in upper plot) and XENON100 
$2011-2012$~\cite{xenon} (dashed line in lower plot). The results of a grid scan of the CMSSM and MSSM-7 are included 
(dark and light grey shaded areas respectively) for the sake of comparison.}
\label{sdsilimitcmssmfig}
\end{center}
\end{figure}

\noindent where $\rm E_{\nu}^{th}=10$~GeV is the energy threshold for neutrino detection in ANTARES, $\rm M_{WIMP}$ 
is the WIMP mass, $\rm dN_{\nu,\bar{\nu}}/dE_{\nu,\bar{\nu}}$ is the energy spectrum of the (anti-)neutrinos at the surface 
of the Earth as shown in Figure~\ref{aeffmwimpfig} (top), and $\rm A_{eff}(E_{\nu,\bar{\nu}})$ is the effective area of 
ANTARES as a function of the (anti-)neutrino energy for tracks coming from the direction of the Sun below the 
horizon. Due to their different cross-sections, the effective areas for neutrinos and anti-neutrinos are slightly 
different and therefore are considered separately. In addition, the fluxes of muon neutrinos and anti-neutrinos 
from the Sun are different and are convoluted with their respective efficiencies.  

An example of an averaged effective area $\rm \bar{A}_{eff}(M_{\rm WIMP})$ for this analysis is shown in Figure~\ref{aeffmwimpfig} 
(bottom) for ($\rm Q_{cut}$,$\rm \Psi_{cut}$) = ($1.4$, $3^{\circ}$) with the visibility of the Sun taken into account, and a 
detector in a 12 line configuration. Whilst the values for each configuration of the detector are detailed in 
Tables~\ref{tab:aeffresultsone} and~\ref{tab:aeffresultstwo} for optimised ($\rm Q_{cut}$,$\rm \Psi_{cut}$) 
(see section~\ref{optimisation}). The corresponding $\rm \bar{A}_{eff}(M_{\rm WIMP})$ distribution of the $W^{+}W^{-}$ 
channel is kinematically allowed for $M_{\rm WIMP} > M_{W} = 80.4$~GeV~\cite{pdg2012}. Note that even though the sensitivity 
$\rm \bar{A}_{eff}(M_{\rm WIMP})$ decreases rapidly with a decreasing WIMP mass, the low mass region, 50~GeV$ < M_{\rm WIMP} < $100~GeV, 
can still be probed.

\bigskip
The cut optimisation procedure provides a pair of optimised values, $\rm Q$ and $\rm \Psi$, for each 
mass of the WIMP and for each studied channel. A value of $\rm Q_{cut} = 1.4$ is found optimum for all considered masses 
and channels. The distribution of the optimal angular separation around the Sun, $\rm \Psi_{cut}$, as a function of the 
WIMP mass is shown in Figure~\ref{psifig}. As the $b\bar{b}$ channel has a softer energy spectrum, $\rm \Psi_{cut}$ 
is larger for this channel. For all the channels, $\rm \Psi_{cut}$ is larger in the low mass regime because of a 
worse angular resolution at low energy ($\rm E_\nu < 100$ GeV). After the optimised $\rm Q_{cut}$ and $\rm \Psi_{cut}$ 
are fixed, the data sample is unblinded.

\section{Results and discussion}
\label{results}

Figure~\ref{databkgfig} shows the distribution of the angular separation between the events and the Sun's 
direction obtained after applying the selection criteria on the zenith angle, the minimum number of hits 
and lines, and a $\rm Q_{cut} = 1.4$. A total of $27$ events are found within a $20^{\circ}$ angular 
separation. No statistically significant excess is observed above the scrambled background in the Sun's 
direction.

Using the values for the cuts obtained in the optimisation procedure, $90$\% CL limits on the 
$\rm \nu_{\mu}+\bar{\nu}_{\mu}$ flux, $\rm \Phi_{\nu_{\mu}+\bar{\nu}_\mu}$, can be computed from the data
according to Equation~\ref{mrfeq}, where the $\rm \bar{\mu}^{90\%}$ average $90$\% CL upper limit is replaced 
by the upper limit at $90$\% CL, $\rm \mu^{90\%}$, on the number of observed events. The corresponding limits 
are presented in Figure~\ref{phinuphimulimitfig} (top) for the three representative self-annihilation 
channels $b\bar{b}$, $W^{+}W^{-}$ and $\tau^{+}\tau^{-}$. Given its soft energy spectrum (see 
Figure~\ref{aeffmwimpfig} (top)), the channel $b\bar{b}$ yields the weakest limit, while the others ($W^{+}W^{-}$, 
$\tau^{+}\tau^{-}$) are the most stringent. 

The corresponding limits on the muon flux are calculated using a conversion factor between the neutrino 
and the muon fluxes ($\rm \Phi_{\mu}=\Gamma_{\nu\rightarrow\mu} \times \Phi_{\nu_{\mu}+\bar{\nu}_\mu}$) 
computed using the package DarkSUSY~\cite{gondolo}. Figure~\ref{phinuphimulimitfig} (bottom) shows the 
90\% CL muon flux limits, $\rm \Phi_{\mu}$, for the channels $b\bar{b}$, $W^{+}W^{-}$ and 
$\tau^{+}\tau^{-}$. The latest results from Baksan~\cite{baksan}, Super-Kamiokande~\cite{superk} and 
IceCube-$79$~\cite{icecube} are also shown for comparison. 

\bigskip
Assuming equilibrium between the WIMP capture and self-annihilation rates 
in the Sun, the limits on the spin-dependent (SD) and the spin-independent (SI) 
WIMP-proton scattering cross-sections are derived for the case in which one 
or the other is dominant.

The Sun is considered to be free in the galactic halo~\cite{sivertsson}. A local dark matter density of $0.3$ GeV/cm$^{3}$ 
and a Maxwellian velocity distribution of the WIMP with a RMS velocity of $270$ km/s are assumed~\cite{pdg2012}, 
and no additional dark matter disk that could enhance the local dark matter density is considered 
(see Reference~\cite{edsjosdcs} for a discussion).

The $90$\% CL limits for the SD, $\rm \sigma_{p,SD}$, and SI, $\rm \sigma_{p,SI}$, WIMP-proton 
cross-sections derived for the signal channels $b\bar{b}$, $W^{+}W^{-}$ and $\tau^{+}\tau^{-}$ 
are presented in Figure~\ref{sdsilimitcmssmfig}. The latest results from Baksan~\cite{baksan}, 
Super-Kamiokande~\cite{superk} and IceCube-$79$~\cite{icecube} together with the latest and the most 
stringent limits from the direct search experiments SIMPLE~\cite{simple}, COUPP~\cite{coupp} and 
XENON100~\cite{xenon} are shown. The allowed parameter space from the CMSSM and MSSM-7 models 
according to the results from an adaptative grid scan performed with DarkSUSY are also shown. 
For CMSSM and MSSM-7, their free parameters are limited as shown in Table~\ref{tab:cmssmsimlimits}. 
All the limits presented in Figure~\ref{sdsilimitcmssmfig} are computed 
with a muon energy threshold at $\rm E_\mu = 1$ GeV. For this figure the shaded regions show a 
grid scan of the model parameter space, taking into account the latest constraints for various 
observables from accelerator-based experiments shown in Table~\ref{tab:aclimits}, in particular 
the results on the Higgs boson mass from ATLAS and CMS, $\rm M_{h} = 125 \pm 2$ GeV~\cite{buchmeller}, and 
the latest limit on the SI WIMP-proton scattering cross-section by XENON100~\cite{xenon}. 
A relatively loose constraint on the neutralino relic density $\rm 0 < \Omega_{CDM}h^{2} < 0.1232$~\cite{wmap7yr} is used 
to take into account the existence of other possible types of dark matter particles. 	

All the results are summarised in Tables~\ref{tab:qresultsone} and~\ref{tab:qresultstwo}, where for each WIMP mass and channel 
the values of the optimised angular separation, the average 90\% CL upper limit computed from the background 
without signal expectation, the 90\% CL upper limit on the number of observed events, the total averaged effective area and the 
90\% CL upper limits are presented. Systematic uncertainties are taken into account and included in the evaluation of the limits using 
the {\tt Pole} software following the approach detailed in Reference~\cite{conrad}. The total systematic uncertainty on the detector 
efficiency is around 20\% and comes mainly from the uncertainties on the average quantum efficiency of the PMTs as well as the angular 
acceptance and the sea water absorption length ($\pm 10$\% for all of them). The detailed uncertainties study is described in 
Reference~\cite{timing}. This total systematic uncertainty translates into a degradation of the upper limit between 3\% and 6\%, 
depending on the WIMP mass. 

\bigskip
The neutrino flux due to WIMP annihilation in the Sun is highly dependent on the capture 
rate of WIMPs in the core of the Sun, which in turn is dominated by the SD WIMP-proton 
cross-section. This makes these indirect searches better compared to direct search 
experiments.  This is not the case for the SI WIMP-proton cross-section, where the limits 
coming from direct search experiments like XENON100 are better thanks to their target			     
materials. 

Using the first two years of data recorded by the ANTARES neutrino telescope, an indirect search for dark matter 
towards the Sun has been performed. The observed number of neutrino events in the Sun's direction is compatible 
with the expectation from the atmospheric backgrounds. The derived limits are comparable with those obtained by 
other neutrino observatories and are more stringent than those obtained by direct search experiments for the 
spin-dependent WIMP-proton scattering cross-section thanks to the hard channels ($W^{+}W^{-}$, 
$\tau^{+}\tau^{-}$). The present ANTARES limits already begin to constrain the parameter 
spaces of the MSSM-7 model.  					     

\section*{Acknowledgments}
The authors acknowledge the financial support of the funding agencies:
Centre National de la Recherche Scientifique (CNRS), Commissariat \`a
l'\'ene\-gie atomique et aux \'energies alternatives (CEA), Agence
National de la Recherche (ANR), Commission Europ\'eenne (FEDER fund
and Marie Curie Program), R\'egion Alsace (contrat CPER), R\'egion
Provence-Alpes-C\^ote d'Azur, D\'e\-par\-tement du Var and Ville de La
Seyne-sur-Mer, France; Bundesministerium f\"ur Bildung und Forschung
(BMBF), Germany; Istituto Nazionale di Fisica Nucleare (INFN), Italy;
Stichting voor Fundamenteel Onderzoek der Materie (FOM), Nederlandse
organisatie voor Wetenschappelijk Onderzoek (NWO), the Netherlands;
Council of the President of the Russian Federation for young
scientists and leading scientific schools supporting grants, Russia;
National Authority for Scientific Research (ANCS), Romania; Ministerio
de Ciencia e Innovaci\'on (MICINN), Prometeo of Generalitat Valenciana
and MultiDark, Spain; Agence de l'Oriental and CNRST, Morocco. We also
acknowledge the technical support of Ifremer, AIM and Foselev Marine
for the sea operation and the CC-IN2P3 for the computing facilities.






\begin{thebibliography}{00}


\bibitem{darkmatter}G. Bertone, D. Hooper, J. Silk, \textit{Particle Dark Matter: Evidence, Candidates and Constraints}, Phys. Rept., 2005, {\bf 405}: pp. 279-390.

\bibitem{pdg2012}J. Beringer et al., Particle Data Group, Phys. Rev. D {\bf 86}, 010001 (2012).

\bibitem{galclust}F. Zwicky, \textit{Die Rotverschiebung von extragalaktischen Nebeln}, Helv. Phys. Acta {\bf 6}, 110 (1933).

\bibitem{galrot}W. J. de Blok, S. S. McGaugh, A. Bosma and V. C. Rubin., \textit{Mass density profiles of LSB galaxies}, Astrophys. J. {552} (2001) L23-L26. 

\bibitem{bulclust}D. Clowe et al., \textit{A direct empirical proof of the existence of dark matter}, Astrophys. J. {648} (2006) L109-L113. 

\bibitem{wmap7yr}E. Komatsu et al., WMAP Collaboration, \textit{Seven-Year Wilkinson Microwave Anisotropy Probe (WMAP) Observations: Cosmological Interpretation}, 
Astrophys. J. Suppl. {192} (2011) 18. 


\bibitem{lsp0}J.R. Ellis et al., \textit{Supersymmetric relics from the big bang}, Nucl. Phys. B {\bf 238}, 453 (1984).

\bibitem{lsp1}H. Goldberg et al., \textit{Constraint on the Photino Mass from Cosmology}, Phys. Rev. Lett. {\bf 50}, 1419 (1983).

\bibitem{CMSSM}J. Ellis et al., \textit{Neutrino Fluxes from CMSSM LSP Annihilations in the Sun}, Phys. Rev. D {\bf 81}, 085004 (2010). 

\bibitem{MSSM7}L. Bergstr\"{o}m and P. Gondolo, \textit{Limits on direct detection of neutralino dark matter from $\rm b \rightarrow s\gamma$ decays}, 
Astropart. Phys. {\bf 5} (1996) 263-278.

\bibitem{indirectdm}A. Gould, \textit{Direct And Indirect Capture Of Wimps By The Earth}, Astrophys. J. {\bf 328}, 919 (1988); 
T. K. Gaisser, G. Steigman, and S. Tilav, \textit{Limits on Cold Dark Matter Candidates from Deep Underground Detectors}, Phys. Rev. D {\bf 34}, 2206 (1986); 
J. Silk et al., \textit{The Photino, the Sun and High-Energy Neutrinos}, Phys. Rev. Lett. {\bf 55}, 257 (1985); 
W. H. Press and D. N. Spergel, \textit{Capture by the sun of a galactic population of weakly interacting, massive particles}, Astrophys. J. {\bf 296}, 679 (1985).

\bibitem{dmsun}A. Gould, \textit{Resonant enhancements in weakly interacting massive particle capture by the earth}, Astrophys. J. {\bf 321}, 571 (1987).


\bibitem{antares} M. Ageron et al., ANTARES Collaboration, \textit{ANTARES: the first undersea neutrino telescope}, 
Nucl. Inst. and Meth. in Phys. Res. A {\bf 656} (2011) 11-38. 

\bibitem{OM}  P. Amram, ANTARES Collaboration, \textit{The ANTARES optical module}, 
Nucl. Inst. and Meth. in Phys. Res. A {\bf 484} (2002) 369.

\bibitem{PMT} J.A. Aguilar et al., ANTARES Collaboration, \textit{Study of large hemispherical photomultiplier tubes for the ANTARES neutrino telescope}, 
Nucl. Inst. and Meth. in Phys. Res. A {\bf 555} (2005) 132.

\bibitem{frontend} J.A. Aguilar et al., ANTARES Collaboration, \textit{Performance of the front-end electronics of the ANTARES Neutrino Telescope}, 
Nucl. Inst. and Meth. in Phys. Res. A {\bf 622} (2010) 59.

\bibitem{DAQ} J.A. Aguilar et al., ANTARES Collaboration, \textit{The data acquisition system for the ANTARES neutrino telescope}, 
Nucl. Inst. and Meth. in Phys. Res. A {\bf 570} (2007) 107.

\bibitem{biofouling} P. Amram, ANTARES Collaboration, \textit{Sedimentation and Fouling of Optical Surfaces at the ANTARES Site}, 
Astropart. Phys. {\bf 19} (2003) 253.

\bibitem{amadeus} J.A. Aguilar et al., ANTARES Collaboration, \textit{AMADEUS - The Acoustic Neutrino Detection Test System of the ANTARES Deep-Sea Neutrino Telescope}, 
Nucl. Inst. and Meth. in Phys. Res. A {\bf 626-627} (2011) 128.

\bibitem{instrumentation}  J.A. Aguilar et al., \textit{First results of the Instrumentation Line for the deep-sea ANTARES neutrino telescope}, 
ANTARES Collaboration, Astropart. Phys. {\bf 26} (2006) 314.

\bibitem{alignment} S. Adri\'an-Mart\'{\i}nez et al., \textit{The positioning system of the ANTARES Neutrino Telescope}, ANTARES Collaboration, JINST {\bf 7} (2012) T08002.

\bibitem{OBs} M. Ageron et al., ANTARES Collaboration, \textit{The Antares optical beacon system}, Nucl. Inst. and Meth. in Phys. Res. A {\bf 578} (2007) 498.

\bibitem{timing} J.A. Aguilar et al., ANTARES Collaboration, \textit{Time Calibration of the ANTARES neutrino Telescope}, Astropart. Phys. {\bf 34} (2011) 539.

\bibitem{diffuse} J.A. Aguilar et al., ANTARES Collaboration, \textit{Search for a diffuse flux of high energy $\rm n_{\mu}$ with the ANTARES neutrino telescope}, 
Phys. Lett, {\bf B696} (2011) 16.

\bibitem{ps}  S. Adri\'an-Mart\'{\i}nez et al., ANTARES Collaboration, \textit{First search for point sources of high energy cosmic neutrinos with the ANTARES neutrino telescope}, 
Ap. J. Letter {\bf 743} (2011) L14.

\bibitem{monopole}  S. Adri\'an-Mart\'{\i}nez et al., ANTARES Collaboration, \textit{Search for relativistic magnetic monopoles with the ANTARES neutrino telescope}, 
Astropart. Phys. {\bf 35} (2012) 634.

\bibitem{bbfit} J.A. Aguilar et al., ANTARES Collaboration, \textit{A fast algorithm for muon track reconstruction and its application to the ANTARES neutrino Telescope}, 
Astropart. Phys. {\bf 34} (2011) 652.


\bibitem{wimpsim}J. Edsj\"{o}, http://www.physto.se/~edsjo/wimpsim/.

\bibitem{abbasi}R. Abbasi et al., IceCube Collaboration, \textit{Limits on a Muon Flux from Neutralino Annihilations in the Sun with the IceCube 22-String Detector}, 
Phys. Rev. Lett. {\bf 102}, 201302 (2009).
\bibitem{boliev}M. M. Boliev et al., Baksan Collaboration, \textit{Search for supersymmetric dark matter with Baksan Underground telescope}, 
Nucl. Phys. B, Proc. Suppl. {\bf 48}, 83 (1996).
\bibitem{ambrosio}M. Ambrosio et al., MACRO Collaboration, \textit{Limits on Dark Matter Wimps using Upward Going Muons in the MACRO Detector}, 
Phys. Rev. D {\bf 60}, 082002 (1999).
\bibitem{desai}S. Desai et al., Super-Kamiokande Collaboration, \textit{Search for dark matter wimps using upward through-going muons in Super-Kamiokande}, 
Phys. Rev. D {\bf 70}, 083523 (2004).
\bibitem{ackermann}M. Ackermann et al., AMANDA Collaboration, \textit{Limits to the muon flux from neutralino annihilations in the Sun with the AMANDA detector}, 
Astropart. Phys. {\bf 24} (2006) 459.


\bibitem{corsika}D. Heck et al., Report FZKA 6019 (1998), Forschungszentrum Karlsruhe; 
D. Heck and J. Knapp, Report FZKA 6097 (1998), Forschungszentrum Karlsruhe.

\bibitem{bartol}G. Barr et al., \textit{Flux of atmospheric neutrinos}, Phys. Rev. D {\bf 39}, 3532 (1989); 
V. Agrawal et al., \textit{Atmospheric neutrino flux above 1 GeV}, Phys. Rev. D {\bf 53}, 1314 (1996).

\bibitem{transmission} J.A. Aguilar et al., ANTARES Collaboration, \textit{Transmission of light in deep sea water at the site of the Antares neutrino telescope}, 
Astropart. Phys. {\bf 23} (2005) 131.

\bibitem{hettlage}C. Hettlage et al., \textit{The sun as a high energy neutrino source}, Astropart.Phys. {\bf 13} (2000) 45-50.

\bibitem{mrf}G.C. Hill, K. Rawlins, \textit{Unbiased cut selection for optimal upper limits in neutrino detectors: the model rejection potential technique}, 
Astropart. Phys. {\bf 19} (2003) 393-402. 
\bibitem{feldmancousins}G.J. Feldman, R.D. Cousins, \textit{A Unified Approach to the Classical Statistical Analysis of Small Signals}, 
Phys. Rev. {\bf D 57}, 3873-3889 (1998). 



\bibitem{gondolo}P. Gondolo et al., \textit{DarkSUSY: Computing Supersymmetric Dark Matter Properties Numerically}, J. Cosm. and Astropart. Phys., JCAP07, 008 (2004).


\bibitem{baksan}M.M. Boliev et al., Baksan Collaboration, 
\textit{Search for muon signal from dark matter annihilations in the Sun with the Baksan Underground Scintillator Telescope for 24.12 years} [astro-ph/1301.1138].

\bibitem{superk}T. Tanaka et al., Super-Kamiokande Collaboration, 
\textit{An Indirect Search for WIMPs in the Sun using 3109.6 days of upward-going muons in Super-Kamiokande}, 
Astrophys. J. {\bf 742}, 78 (2011).

\bibitem{icecube}M.G. Aartsen et al., IceCube Collaboration, \textit{Search for dark matter annihilations in the Sun with the 79-string IceCube detector}, [astro-ph/1212.4097].



\bibitem{sivertsson}S. Sivertsson and J. Edsj\"{o}, \textit{WIMP diffusion in the Solar System including Solar WIMP-nucleon scattering}, 
Phys. Rev. {\bf D 85}, 123514 (2012).

\bibitem{edsjosdcs}G. Wikstr\"{o}m and J. Edsj\"{o}, \textit{Limits on the WIMP-nucleon scattering cross-section from neutrino telescopes}, 
J. Cosm. and Astropart. Phys., JCAP04, 009 (2009).

\bibitem{simple}M. Felizardo et al., SIMPLE Collaboration, \textit{Final Analysis and Results of the Phase II SIMPLE Dark Matter Search}, 
Phys. Rev. Lett. {\bf 108}, 201302 (2012). 
\bibitem{coupp}E. Behnke et al., COUPP Collaboration, \textit{First dark matter search results from a 4-kg CF$_3$I bubble chamber operated in a deep underground site}, 
Phys. Rev. D {\bf 86}, 052001 (2012). 

\bibitem{xenon}E. Aprile et al., XENON Collaboration, \textit{Dark Matter Results from 225 Live Days of XENON100 Data}, [astro-ph/1207.5988].

\bibitem{buchmeller}O. Buchmueller et al., \textit{The CMSSM and NUHM1 in Light of 7 TeV LHC, B\_s to mu+mu- and XENON100 Data}, [hep-ph/1207.7315].

\bibitem{conrad}F. Tegenfeldt, J. Conrad, \textit{On Bayesian Treatment of Systematic Uncertainties In Confidence Interval Calculation}, 
Nucl. Inst. and Meth. in Phys. Res. A {\bf 539} (2005) 407-413; 
J. Conrad et al., \textit{Including Systematic Uncertainties in Confidence Interval Construction for Poisson Statistics}, Phys. Rev. D {\bf 67}, 012002 (2003); 
J. Conrad, \textit{Discovery and Upper Limits in Search for Exotic Physics with Neutrino Telescopes}, [astro-ph/0612082].

\end{thebibliography}



\newpage

\begin{table*}[!b]
\begin{center}
\begin{footnotesize}
\begin{tabular}{cccccc}
\hline
\hline
&&&&& \\[-0.1in]
$M_{\rm WIMP}$ & Channel & $\rm \bar{A}_{eff}^{5L}(M_{\rm WIMP})$ & $\rm \bar{A}_{eff}^{9L}(M_{\rm WIMP})$ & $\rm \bar{A}_{eff}^{10L}(M_{\rm WIMP})$ & $\rm \bar{A}_{eff}^{12L}(M_{\rm WIMP})$ \\
(GeV) & & (m$^2$) & (m$^2$) & (m$^2$) & (m$^2$) \\
&&&&& \\[-0.1in]
\hline
&&&&& \\[-0.1in]
$50$ & $b\bar{b}$ & $3.5\times 10^{-10}$ & $7.3\times 10^{-10}$ & $1.2\times 10^{-9}$ & $1.6\times 10^{-9}$ \\
$$ & $\tau\bar{\tau}$ & $5.5\times 10^{-8}$ & $1.0\times 10^{-7}$ & $1.3\times 10^{-7}$ & $1.6\times 10^{-7}$ \\
&&&&& \\[-0.1in]
\hline
&&&&& \\[-0.1in]
$80.3$ & $b\bar{b}$ & $5.4\times 10^{-9}$ & $9.9\times 10^{-9}$ & $1.3\times 10^{-8}$ & $1.5\times 10^{-8}$ \\
$$ & $W^{+}W^{-}$ & $2.7\times 10^{-7}$ & $4.7\times 10^{-7}$ & $5.9\times 10^{-7}$ & $9.0\times 10^{-7}$ \\
$$ & $\tau\bar{\tau}$ & $3.7\times 10^{-7}$ & $6.8\times 10^{-7}$ & $9.6\times 10^{-7}$ & $1.4\times 10^{-6}$ \\
&&&&& \\[-0.1in]
\hline
&&&&& \\[-0.1in]
$100$ & $b\bar{b}$ & $1.4\times 10^{-8}$ & $2.5\times 10^{-8}$ & $3.4\times 10^{-8}$ & $4.4\times 10^{-8}$ \\
$$ & $W^{+}W^{-}$ & $8.2\times 10^{-7}$ & $1.6\times 10^{-6}$ & $2.2\times 10^{-6}$ & $3.2\times 10^{-6}$ \\
$$ & $\tau\bar{\tau}$ & $7.5\times 10^{-7}$ & $1.4\times 10^{-6}$ & $2.1\times 10^{-6}$ & $3.0\times 10^{-6}$ \\
&&&&& \\[-0.1in]
\hline
&&&&& \\[-0.1in]
$150$ & $b\bar{b}$ & $5.5\times 10^{-8}$ & $9.9\times 10^{-8}$ & $1.4\times 10^{-7}$ & $2.0\times 10^{-7}$ \\
$$ & $W^{+}W^{-}$ & $2.8\times 10^{-6}$ & $4.9\times 10^{-6}$ & $8.4\times 10^{-6}$ & $1.2\times 10^{-5}$ \\
$$ & $\tau\bar{\tau}$ & $2.2\times 10^{-6}$ & $3.9\times 10^{-6}$ & $6.5\times 10^{-6}$ & $9.2\times 10^{-6}$ \\
&&&&& \\[-0.1in]
\hline
&&&&& \\[-0.1in]
$176$ & $b\bar{b}$ & $8.7\times 10^{-8}$ & $1.6\times 10^{-7}$ & $2.3\times 10^{-7}$ & $3.2\times 10^{-7}$ \\
$$ & $W^{+}W^{-}$ & $4.2\times 10^{-6}$ & $7.5\times 10^{-6}$ & $1.3\times 10^{-5}$ & $1.8\times 10^{-5}$ \\
$$ & $\tau\bar{\tau}$ & $3.2\times 10^{-6}$ & $5.8\times 10^{-6}$ & $9.8\times 10^{-6}$ & $1.4\times 10^{-5}$ \\
&&&&& \\[-0.1in]
\hline
&&&&& \\[-0.1in]
$200$ & $b\bar{b}$ & $1.2\times 10^{-7}$ & $2.2\times 10^{-7}$ & $3.2\times 10^{-7}$ & $4.6\times 10^{-7}$ \\
$$ & $W^{+}W^{-}$ & $5.3\times 10^{-6}$ & $9.4\times 10^{-6}$ & $1.6\times 10^{-5}$ & $2.2\times 10^{-5}$ \\
$$ & $\tau\bar{\tau}$ & $4.3\times 10^{-6}$ & $7.7\times 10^{-6}$ & $1.3\times 10^{-5}$ & $1.8\times 10^{-5}$ \\
&&&&& \\[-0.1in]
\hline
&&&&& \\[-0.1in]
$250$ & $b\bar{b}$ & $2.1\times 10^{-7}$ & $3.9\times 10^{-7}$ & $5.9\times 10^{-7}$ & $8.4\times 10^{-7}$ \\
$$ & $W^{+}W^{-}$ & $7.9\times 10^{-6}$ & $1.3\times 10^{-5}$ & $2.5\times 10^{-5}$ & $3.3\times 10^{-5}$ \\
$$ & $\tau\bar{\tau}$ & $6.5\times 10^{-6}$ & $1.1\times 10^{-5}$ & $2.0\times 10^{-5}$ & $2.7\times 10^{-5}$ \\
&&&&& \\[-0.1in]
\hline
&&&&& \\[-0.1in]
$350$ & $b\bar{b}$ & $4.3\times 10^{-7}$ & $7.9\times 10^{-7}$ & $1.2\times 10^{-6}$ & $1.7\times 10^{-6}$ \\
$$ & $W^{+}W^{-}$ & $1.3\times 10^{-5}$ & $2.2\times 10^{-5}$ & $4.0\times 10^{-5}$ & $5.4\times 10^{-5}$ \\
$$ & $\tau\bar{\tau}$ & $1.2\times 10^{-5}$ & $2.0\times 10^{-5}$ & $3.6\times 10^{-5}$ & $4.8\times 10^{-5}$ \\
&&&&& \\[-0.1in]
\hline
\hline
\end{tabular}
\caption{Detailed numerical values of the averaged effective areas $\rm \bar{A}_{eff}^{i}(M_{\rm WIMP})$ for the signal of 
WIMP self-annihilation inside the Sun, $b\bar{b}$, $W^{+}W^{-}$ and $\tau^{+}\tau^{-}$ channels. The 5, 9, 10 and 12 line 
configurations (i index) with ($\rm Q_{cut}$,$\rm \Psi_{cut}$) after optimisation (see section~\ref{optimisation}) are considered.
The total averaged effective area $\rm \bar{A}_{eff}(M_{\rm WIMP}) = \sum_{i} \bar{A}_{eff}^{i}(M_{WIMP}) \times T_{eff}^{i}$ 
(see Equation~\ref{aeffeq}) is reported in Tables~\ref{tab:qresultsone} and~\ref{tab:qresultstwo}. Results for $\rm M_{\rm WIMP}>350$ GeV 
are available in Table~\ref{tab:aeffresultstwo}. 
\label{tab:aeffresultsone}}
\end{footnotesize}
\end{center}
\end{table*}

\begin{table*}[!b]
\begin{center}
\begin{footnotesize}
\begin{tabular}{cccccc}
\hline
\hline
&&&&& \\[-0.1in]
$M_{\rm WIMP}$ & Channel & $\rm \bar{A}_{eff}^{5L}(M_{\rm WIMP})$ & $\rm \bar{A}_{eff}^{9L}(M_{\rm WIMP})$ & $\rm \bar{A}_{eff}^{10L}(M_{\rm WIMP})$ & $\rm \bar{A}_{eff}^{12L}(M_{\rm WIMP})$ \\
(GeV) & & (m$^2$) & (m$^2$) & (m$^2$) & (m$^2$) \\
&&&&& \\[-0.1in]
\hline
&&&&& \\[-0.1in]
$500$ & $b\bar{b}$ & $7.5\times 10^{-7}$ & $1.3\times 10^{-6}$ & $2.2\times 10^{-6}$ & $3.1\times 10^{-6}$ \\
$$ & $W^{+}W^{-}$ & $1.9\times 10^{-5}$ & $3.4\times 10^{-5}$ & $5.9\times 10^{-5}$ & $8.0\times 10^{-5}$ \\
$$ & $\tau\bar{\tau}$ & $1.9\times 10^{-5}$ & $3.2\times 10^{-5}$ & $5.7\times 10^{-5}$ & $7.6\times 10^{-5}$ \\
&&&&& \\[-0.1in]
\hline
&&&&& \\[-0.1in]
$750$ & $b\bar{b}$ & $1.3\times 10^{-6}$ & $2.2\times 10^{-6}$ & $3.8\times 10^{-6}$ & $5.2\times 10^{-6}$ \\
$$ & $W^{+}W^{-}$ & $2.6\times 10^{-5}$ & $4.5\times 10^{-5}$ & $7.8\times 10^{-5}$ & $1.0\times 10^{-4}$ \\
$$ & $\tau\bar{\tau}$ & $2.7\times 10^{-5}$ & $4.6\times 10^{-5}$ & $8.1\times 10^{-5}$ & $1.1\times 10^{-4}$ \\
&&&&& \\[-0.1in]
\hline
&&&&& \\[-0.1in]
$1000$ & $b\bar{b}$ & $1.7\times 10^{-6}$ & $3.0\times 10^{-6}$ & $5.2\times 10^{-6}$ & $7.1\times 10^{-6}$ \\
$$ & $W^{+}W^{-}$ & $2.7\times 10^{-5}$ & $4.8\times 10^{-5}$ & $8.3\times 10^{-5}$ & $1.1\times 10^{-4}$ \\
$$ & $\tau\bar{\tau}$ & $3.2\times 10^{-5}$ & $5.6\times 10^{-5}$ & $9.8\times 10^{-5}$ & $1.3\times 10^{-4}$ \\
&&&&& \\[-0.1in]
\hline
&&&&& \\[-0.1in]
$1500$ & $b\bar{b}$ & $2.3\times 10^{-6}$ & $4.0\times 10^{-6}$ & $7.0\times 10^{-6}$ & $9.4\times 10^{-6}$ \\
$$ & $W^{+}W^{-}$ & $2.7\times 10^{-5}$ & $4.8\times 10^{-5}$ & $8.4\times 10^{-5}$ & $1.1\times 10^{-4}$ \\
$$ & $\tau\bar{\tau}$ & $3.3\times 10^{-5}$ & $5.9\times 10^{-5}$ & $1.0\times 10^{-4}$ & $1.3\times 10^{-4}$ \\
&&&&& \\[-0.1in]
\hline
&&&&& \\[-0.1in]
$2000$ & $b\bar{b}$ & $2.7\times 10^{-6}$ & $4.5\times 10^{-6}$ & $8.2\times 10^{-6}$ & $1.1\times 10^{-5}$ \\
$$ & $W^{+}W^{-}$ & $2.6\times 10^{-5}$ & $4.6\times 10^{-5}$ & $8.1\times 10^{-5}$ & $1.1\times 10^{-4}$ \\
$$ & $\tau\bar{\tau}$ & $3.3\times 10^{-5}$ & $5.8\times 10^{-5}$ & $1.0\times 10^{-4}$ & $1.4\times 10^{-4}$ \\
&&&&& \\[-0.1in]
\hline
&&&&& \\[-0.1in]
$3000$ & $b\bar{b}$ & $3.0\times 10^{-6}$ & $5.1\times 10^{-6}$ & $9.3\times 10^{-6}$ & $1.2\times 10^{-5}$ \\
$$ & $W^{+}W^{-}$ & $2.4\times 10^{-5}$ & $4.2\times 10^{-5}$ & $7.4\times 10^{-5}$ & $9.7\times 10^{-5}$ \\
$$ & $\tau\bar{\tau}$ & $3.0\times 10^{-5}$ & $5.4\times 10^{-5}$ & $9.5\times 10^{-5}$ & $1.2\times 10^{-4}$ \\
&&&&& \\[-0.1in]
\hline
&&&&& \\[-0.1in]
$5000$ & $b\bar{b}$ & $3.5\times 10^{-6}$ & $5.8\times 10^{-6}$ & $1.0\times 10^{-5}$ & $1.3\times 10^{-5}$ \\
$$ & $W^{+}W^{-}$ & $2.2\times 10^{-5}$ & $3.9\times 10^{-5}$ & $6.6\times 10^{-5}$ & $8.6\times 10^{-5}$ \\
$$ & $\tau\bar{\tau}$ & $2.7\times 10^{-5}$ & $4.8\times 10^{-5}$ & $8.0\times 10^{-5}$ & $1.0\times 10^{-4}$ \\
&&&&& \\[-0.1in]
\hline
&&&&& \\[-0.1in]
$10000$ & $b\bar{b}$ & $3.5\times 10^{-6}$ & $5.5\times 10^{-6}$ & $9.9\times 10^{-6}$ & $1.3\times 10^{-5}$ \\
$$ & $W^{+}W^{-}$ & $1.6\times 10^{-5}$ & $2.8\times 10^{-5}$ & $4.9\times 10^{-5}$ & $6.5\times 10^{-5}$ \\
$$ & $\tau\bar{\tau}$ & $1.8\times 10^{-5}$ & $3.1\times 10^{-5}$ & $5.4\times 10^{-5}$ & $7.2\times 10^{-5}$ \\
&&&&& \\[-0.1in]
\hline
\hline
\end{tabular}
\caption{Detailed numerical values of the averaged effective areas $\rm \bar{A}_{eff}^{i}(M_{\rm WIMP})$ for the signal of 
WIMP self-annihilation inside the Sun, $b\bar{b}$, $W^{+}W^{-}$ and $\tau^{+}\tau^{-}$ channels. The 5, 9, 10 and 12 line 
configurations (i index) with ($\rm Q_{cut}$,$\rm \Psi_{cut}$) after optimisation (see section~\ref{optimisation}) are considered.
The total averaged effective area $\rm \bar{A}_{eff}(M_{\rm WIMP}) = \sum_{i} \bar{A}_{eff}^{i}(M_{WIMP}) \times T_{eff}^{i}$ 
(see Equation~\ref{aeffeq}) is reported in Tables~\ref{tab:qresultsone} and~\ref{tab:qresultstwo}. Results for $\rm M_{\rm WIMP}<500$ GeV 
are available in Table~\ref{tab:aeffresultsone}.
\label{tab:aeffresultstwo}}
\end{footnotesize}
\end{center}
\end{table*}

\clearpage

\begin{table*}[!t]
\begin{center}
\begin{footnotesize}
\begin{tabular}{ccc}
\hline
\hline
&& \\[-0.1in]
Model & Parameter & Range \\
&& \\[-0.1in]
\hline
& Common scalar mass & $50$ GeV$\rm <m_{0}<4$ TeV\\
& Common gaugino mass & $500$ GeV$\rm <m_{1/2}<2.5$ TeV\\
CMSSM & Ratio of vevs of the Higgs fields & $\rm 5<tan(\beta)<62$\\
& Common trilinear coupling & $-5$ TeV$\rm <A_{0}<5$ TeV\\
& Sign of the Higgs mixing & $\rm sgn(\mu) > 0$\\
\hline
\hline
& Higgsino mass term & $-10$ TeV$\rm <\mu<10$ TeV\\
& Gaugino mass term & $-10$ TeV$\rm <M_{2}<10$ TeV\\
MSSM-7 & CP-odd Higgs boson mass & $60$ GeV$\rm <m_{A}<1$ TeV\\
& Trilinear couplings for & $\rm -3m_{0}<A_{b}<3m_{0}$ \\
& the third generation squarks & $\rm -3m_{0}<A_{t}<3m_{0}$\\
\hline
\hline
\end{tabular}
\caption{Range of parameters scanned for the CMSSM and MSSM-7 models.
\label{tab:cmssmsimlimits}}
\end{footnotesize}
\end{center}
\end{table*}

\begin{table*}[!b]
\begin{center}
\begin{footnotesize}
\begin{tabular}{cc}
\hline
\hline
& \\[-0.1in]
Observable & Lower limit ($95$\% CL) \\
& \\[-0.1in]
\hline
$\rm m_{\tilde{\chi}_{1}^{\pm}}$&$>94$ GeV \\
$\rm m_{\tilde{g}}$&$>500$ GeV \\
$\rm m_{\tilde{q}}$&$>1100$ GeV \\
$\rm m_{\tilde{e}_{L}}$&$>107$ GeV \\
$\rm m_{\tilde{e}_{R}}$&$>73$ GeV \\
$\rm m_{\tilde{\mu}_{L,R}}$&$>94$ GeV \\
$\rm m_{\tilde{\tau}_{L,R}}$&$>81.9$ GeV \\
$\rm m_{\tilde{\nu}}$&$>43.7$ GeV \\
$\rm m_{\tilde{\chi}_{1}^{0}}$&$>46$ GeV \\
$\rm m_{\tilde{\chi}_{2}^{0}}$&$>62.4$ GeV \\
$\rm m_{\tilde{\chi}_{3}^{0}}$&$>99.9$ GeV \\
$\rm m_{\tilde{\chi}_{4}^{0}}$&$>116$ GeV \\
$\rm m_{H^{\pm}}$&$>79.3$ GeV \\
$\rm g^{\nu_{l}}$&$>0.502$  \\
\hline
\hline
& \\[-0.1in]
Observable & Value \\
& \\[-0.1in]
\hline
$\rm M_{h}$&$125 \pm 2$ GeV \\
$\rm \delta a_{\mu}^{SUSY}$&$(28.7 \pm 16) \times 10^{-10}$ \\
$\rm BR(\bar{B} \rightarrow X_{s}\gamma)$&$(3.55 \pm 0.84) \times 10^{-4}$ \\
\hline
\hline
\end{tabular}
\caption{Summary of the observables used in the grid scan performed with the package DarkSUSY on the CMSSM and MSSM-7 
free parameter space. {\bf Top:} Observables for which only limits currently exist. The mass of the chargino $\rm \tilde{\chi}_{1}^{\pm}$, 
gluino $\rm \tilde{g}$, squarks $\rm \tilde{q}$, sleptons $\rm \tilde{e}_{L}$, $\rm \tilde{e}_{R}$, $\rm \tilde{\mu}_{L,R}$ and $\rm \tilde{\tau}_{L,R}$, 
sneutrinos $\rm \tilde{\nu}$, neutralinos $\rm \tilde{\chi}_{1}^{0}$ (LSP and dark matter candidate in this analysis), $\rm \tilde{\chi}_{2}^{0}$, 
$\rm \tilde{\chi}_{3}^{0}$ and $\rm \tilde{\chi}_{4}^{0}$, charged Higgs $\rm H^{\pm}$ and the effective neutrino coupling $\rm g^{\nu_{l}}$ from 
invisible Z-decay width~\cite{pdg2012}. {\bf Bottom:} Observables for which a measurement is available. The Higgs boson $h$ mass 
as an averaged result from CMS and ATLAS Collaborations~\cite{buchmeller}, the discrepancy $\rm \delta a_{\mu}^{SUSY}$ between the experimental 
value and the SM prediction of the anomalous magnetic moment of the muon $\rm (g-2)_{\mu}$~\cite{pdg2012} and the branching ratio of the b-hadron 
decay $\rm \bar{B} \rightarrow X_{s}\gamma$~\cite{pdg2012}.
\label{tab:aclimits}}
\end{footnotesize}
\end{center}
\end{table*}

\setlength{\tabcolsep}{3.4pt}

\begin{table*}[!b]
\begin{center}
\begin{footnotesize}
\begin{tabular}{cccccccccc}
\hline
\hline
&&&&&&&&& \\[-0.1in]
$M_{\rm WIMP}$ & Channel & $\rm \Psi_{cut}$ & $\rm \mu^{90\%}$ & $\rm \bar{A}_{eff}(M_{\rm WIMP})$ & $\rm \overline{\Phi}_{\nu_{\mu}+\bar{\nu}_\mu}$ & $\rm \Phi_{\nu_{\mu}+\bar{\nu}_\mu}$ & $\rm \Phi_{\mu}$ & $\rm \sigma_{p,SD}$ & $\rm \sigma_{p,SI}$ \\
(GeV) & & ($^{\circ}$) & & (m$^2$) & (km$^{-2}$/yr) & (km$^{-2}$/yr) & (km$^{-2}$/yr) & (pb) & (pb) \\
&&&&&&&&& \\[-0.1in]

&&&&&&&&& \\[-0.1in]
\hline
&&&&&&&&& \\[-0.1in]
$50$ & $b\bar{b}$ & $8.4$ & $7.5$ & $6.9\times 10^{-10}$ & $7.5\times 10^{15}$ & $1.1\times 10^{16}$ & $2.4\times 10^{5}$ & $7.6\times 10^{-1}$ & $2.9\times 10^{-3}$ \\
$$ & $\tau\bar{\tau}$ & $5.7$ & $5.1$ & $8.2\times 10^{-8}$ & $5.0\times 10^{13}$ & $6.2\times 10^{13}$ & $1.1\times 10^{4}$ & $1.4\times 10^{-3}$ & $5.5\times 10^{-6}$ \\
&&&&&&&&& \\[-0.1in]
\hline
&&&&&&&&& \\[-0.1in]
$80.3$ & $b\bar{b}$ & $5.7$ & $5.1$ & $7.8\times 10^{-9}$ & $5.2\times 10^{14}$ & $6.6\times 10^{14}$ & $2.6\times 10^{4}$ & $1.0\times 10^{-1}$ & $2.7\times 10^{-4}$ \\
$$ & $W^{+}W^{-}$ & $5.7$ & $5.1$ & $4.2\times 10^{-7}$ & $9.7\times 10^{12}$ & $1.2\times 10^{13}$ & $6.0\times 10^{3}$ & $1.8\times 10^{-3}$ & $4.6\times 10^{-6}$ \\
$$ & $\tau\bar{\tau}$ & $5.2$ & $5.5$ & $6.3\times 10^{-7}$ & $6.1\times 10^{12}$ & $8.8\times 10^{12}$ & $3.9\times 10^{3}$ & $4.8\times 10^{-4}$ & $1.3\times 10^{-6}$ \\
&&&&&&&&& \\[-0.1in]
\hline
&&&&&&&&& \\[-0.1in]
$100$ & $b\bar{b}$ & $5.7$ & $5.1$ & $2.2\times 10^{-8}$ & $1.9\times 10^{14}$ & $2.4\times 10^{14}$ & $1.4\times 10^{4}$ & $5.5\times 10^{-2}$ & $1.2\times 10^{-4}$ \\
$$ & $W^{+}W^{-}$ & $5.1$ & $5.6$ & $1.4\times 10^{-6}$ & $2.7\times 10^{12}$ & $3.9\times 10^{12}$ & $3.1\times 10^{3}$ & $8.5\times 10^{-4}$ & $1.9\times 10^{-6}$ \\
$$ & $\tau\bar{\tau}$ & $5.2$ & $5.5$ & $1.3\times 10^{-6}$ & $2.9\times 10^{12}$ & $4.1\times 10^{12}$ & $2.9\times 10^{3}$ & $3.4\times 10^{-4}$ & $7.6\times 10^{-7}$ \\
&&&&&&&&& \\[-0.1in]
\hline
&&&&&&&&& \\[-0.1in]
$150$ & $b\bar{b}$ & $5.2$ & $5.5$ & $9.1\times 10^{-8}$ & $4.3\times 10^{13}$ & $6.1\times 10^{13}$ & $6.9\times 10^{3}$ & $3.1\times 10^{-2}$ & $5.1\times 10^{-5}$ \\
$$ & $W^{+}W^{-}$ & $4.6$ & $5.9$ & $5.1\times 10^{-6}$ & $7.0\times 10^{11}$ & $1.2\times 10^{12}$ & $2.0\times 10^{3}$ & $5.6\times 10^{-4}$ & $9.4\times 10^{-7}$ \\
$$ & $\tau\bar{\tau}$ & $4.6$ & $5.9$ & $4.0\times 10^{-6}$ & $9.0\times 10^{11}$ & $1.5\times 10^{12}$ & $2.1\times 10^{3}$ & $2.7\times 10^{-4}$ & $4.5\times 10^{-7}$ \\
&&&&&&&&& \\[-0.1in]
\hline
&&&&&&&&& \\[-0.1in]
$176$ & $b\bar{b}$ & $5.2$ & $5.5$ & $1.5\times 10^{-7}$ & $2.6\times 10^{13}$ & $3.8\times 10^{13}$ & $5.5\times 10^{3}$ & $2.5\times 10^{-2}$ & $3.8\times 10^{-5}$ \\
$$ & $W^{+}W^{-}$ & $4.6$ & $5.9$ & $7.8\times 10^{-6}$ & $4.6\times 10^{11}$ & $7.6\times 10^{11}$ & $1.8\times 10^{3}$ & $5.0\times 10^{-4}$ & $7.7\times 10^{-7}$ \\
$$ & $\tau\bar{\tau}$ & $4.6$ & $5.9$ & $6.0\times 10^{-6}$ & $6.0\times 10^{11}$ & $9.9\times 10^{11}$ & $1.9\times 10^{3}$ & $2.5\times 10^{-4}$ & $3.8\times 10^{-7}$ \\
&&&&&&&&& \\[-0.1in]
\hline
&&&&&&&&& \\[-0.1in]
$200$ & $b\bar{b}$ & $5.2$ & $5.5$ & $2.1\times 10^{-7}$ & $1.9\times 10^{13}$ & $2.7\times 10^{13}$ & $4.7\times 10^{3}$ & $2.3\times 10^{-2}$ & $3.2\times 10^{-5}$ \\
$$ & $W^{+}W^{-}$ & $4.2$ & $3.1$ & $9.7\times 10^{-6}$ & $3.5\times 10^{11}$ & $3.2\times 10^{11}$ & $8.9\times 10^{2}$ & $2.8\times 10^{-4}$ & $3.9\times 10^{-7}$ \\
$$ & $\tau\bar{\tau}$ & $4.6$ & $5.9$ & $8.0\times 10^{-6}$ & $4.5\times 10^{11}$ & $7.4\times 10^{11}$ & $1.7\times 10^{3}$ & $2.4\times 10^{-4}$ & $3.4\times 10^{-7}$ \\
&&&&&&&&& \\[-0.1in]
\hline
&&&&&&&&& \\[-0.1in]
$250$ & $b\bar{b}$ & $5.2$ & $5.5$ & $3.8\times 10^{-7}$ & $1.0\times 10^{13}$ & $1.5\times 10^{13}$ & $3.6\times 10^{3}$ & $1.9\times 10^{-2}$ & $2.4\times 10^{-5}$ \\
$$ & $W^{+}W^{-}$ & $4.1$ & $3.2$ & $1.4\times 10^{-5}$ & $2.4\times 10^{11}$ & $2.2\times 10^{11}$ & $8.5\times 10^{2}$ & $3.0\times 10^{-4}$ & $3.8\times 10^{-7}$ \\
$$ & $\tau\bar{\tau}$ & $4.2$ & $3.1$ & $1.2\times 10^{-5}$ & $2.9\times 10^{11}$ & $2.6\times 10^{11}$ & $8.4\times 10^{2}$ & $1.3\times 10^{-4}$ & $1.6\times 10^{-7}$ \\
&&&&&&&&& \\[-0.1in]
\hline
&&&&&&&&& \\[-0.1in]
$350$ & $b\bar{b}$ & $5.2$ & $5.5$ & $7.7\times 10^{-7}$ & $5.0\times 10^{12}$ & $7.1\times 10^{12}$ & $2.7\times 10^{3}$ & $1.8\times 10^{-2}$ & $1.9\times 10^{-5}$ \\
$$ & $W^{+}W^{-}$ & $3.8$ & $3.3$ & $2.4\times 10^{-5}$ & $1.4\times 10^{11}$ & $1.4\times 10^{11}$ & $8.0\times 10^{2}$ & $3.9\times 10^{-4}$ & $4.1\times 10^{-7}$ \\
$$ & $\tau\bar{\tau}$ & $4.1$ & $3.2$ & $2.1\times 10^{-5}$ & $1.6\times 10^{11}$ & $1.5\times 10^{11}$ & $7.7\times 10^{2}$ & $1.5\times 10^{-4}$ & $1.6\times 10^{-7}$ \\
\hline
\hline
\end{tabular}
\caption{Results after optimisation and unblinding for the angular separation $\Psi_{cut}$, the $90\%$ CL upper limit on the expected signal $\rm \mu^{90\%}$, the total averaged 
effective area $\rm \bar{A}_{eff}(M_{\rm WIMP}) = \sum_{i} \bar{A}_{eff}^{i}(M_{WIMP}) \times T_{eff}^{i}$ (with $i$ corresponding to a given period of the detector), the 
$90\%$ CL sensitivities on the neutrino+anti-neutrino flux at the Earth $\rm \overline{\Phi}_{\nu_{\mu}+\bar{\nu}_\mu}$, and the $90\%$ CL limits on the neutrino+anti-neutrino 
flux at the Earth $\rm \Phi_{\nu_{\mu}+\bar{\nu}_\mu}$, on the muon flux at the detector $\rm \Phi_{\mu}$ ($E_{\mu} > 1$ GeV), and on the spin-dependent and spin-independent 
WIMP-proton cross-sections $\rm \sigma_{p,SD}$ and $\rm \sigma_{p,SI}$ respectively. Results for $\rm M_{\rm WIMP}>350$ GeV are available in Table~\ref{tab:qresultstwo}.
\label{tab:qresultsone}}
\end{footnotesize}
\end{center}
\end{table*}

\begin{table*}[!b]
\begin{center}
\begin{footnotesize}
\begin{tabular}{cccccccccc}
\hline
\hline
&&&&&&&&& \\[-0.1in]
$M_{\rm WIMP}$ & Channel & $\rm \Psi_{cut}$ & $\rm \mu^{90\%}$ & $\rm \bar{A}_{eff}(M_{\rm WIMP})$ & $\rm \overline{\Phi}_{\nu_{\mu}+\bar{\nu}_\mu}$ & $\rm \Phi_{\nu_{\mu}+\bar{\nu}_\mu}$ & $\rm \Phi_{\mu}$ & $\rm \sigma_{p,SD}$ & $\rm \sigma_{p,SI}$ \\
(GeV) & & ($^{\circ}$) & & (m$^2$) & (km$^{-2}$/yr) & (km$^{-2}$/yr) & (km$^{-2}$/yr) & (pb) & (pb) \\
&&&&&&&&& \\[-0.1in]

&&&&&&&&& \\[-0.1in]
\hline
&&&&&&&&& \\[-0.1in]
$500$ & $b\bar{b}$ & $4.6$ & $5.9$ & $1.4\times 10^{-6}$ & $2.7\times 10^{12}$ & $4.4\times 10^{12}$ & $2.4\times 10^{3}$ & $2.2\times 10^{-2}$ & $2.0\times 10^{-5}$ \\
$$ & $W^{+}W^{-}$ & $3.6$ & $1.7$ & $3.5\times 10^{-5}$ & $9.1\times 10^{10}$ & $4.7\times 10^{10}$ & $3.8\times 10^{2}$ & $2.7\times 10^{-4}$ & $2.5\times 10^{-7}$ \\
$$ & $\tau\bar{\tau}$ & $3.8$ & $3.3$ & $3.4\times 10^{-5}$ & $9.7\times 10^{10}$ & $9.9\times 10^{10}$ & $7.6\times 10^{2}$ & $2.0\times 10^{-4}$ & $1.8\times 10^{-7}$ \\
&&&&&&&&& \\[-0.1in]
\hline
&&&&&&&&& \\[-0.1in]
$750$ & $b\bar{b}$ & $4.6$ & $5.9$ & $2.3\times 10^{-6}$ & $1.6\times 10^{12}$ & $2.6\times 10^{12}$ & $2.1\times 10^{3}$ & $2.9\times 10^{-2}$ & $2.4\times 10^{-5}$ \\
$$ & $W^{+}W^{-}$ & $3.6$ & $1.7$ & $4.6\times 10^{-5}$ & $6.9\times 10^{10}$ & $3.6\times 10^{10}$ & $3.7\times 10^{2}$ & $4.9\times 10^{-4}$ & $4.0\times 10^{-7}$ \\
$$ & $\tau\bar{\tau}$ & $3.6$ & $1.7$ & $4.8\times 10^{-5}$ & $6.7\times 10^{10}$ & $3.4\times 10^{10}$ & $3.7\times 10^{2}$ & $1.6\times 10^{-4}$ & $1.3\times 10^{-7}$ \\
&&&&&&&&& \\[-0.1in]
\hline
&&&&&&&&& \\[-0.1in]
$1000$ & $b\bar{b}$ & $4.2$ & $3.1$ & $3.1\times 10^{-6}$ & $1.1\times 10^{12}$ & $1.0\times 10^{12}$ & $1.0\times 10^{4}$ & $2.0\times 10^{-2}$ & $1.5\times 10^{-5}$ \\
$$ & $W^{+}W^{-}$ & $3.2$ & $1.8$ & $4.9\times 10^{-5}$ & $6.3\times 10^{10}$ & $3.5\times 10^{10}$ & $4.0\times 10^{2}$ & $8.9\times 10^{-4}$ & $6.9\times 10^{-7}$ \\
$$ & $\tau\bar{\tau}$ & $3.6$ & $1.7$ & $5.8\times 10^{-5}$ & $5.5\times 10^{10}$ & $2.9\times 10^{10}$ & $3.6\times 10^{2}$ & $2.3\times 10^{-4}$ & $1.8\times 10^{-7}$ \\
&&&&&&&&& \\[-0.1in]
\hline
&&&&&&&&& \\[-0.1in]
$1500$ & $b\bar{b}$ & $4.1$ & $3.2$ & $4.2\times 10^{-6}$ & $8.2\times 10^{11}$ & $7.7\times 10^{11}$ & $9.7\times 10^{3}$ & $3.4\times 10^{-2}$ & $2.5\times 10^{-5}$ \\
$$ & $W^{+}W^{-}$ & $3.3$ & $1.8$ & $4.9\times 10^{-5}$ & $6.3\times 10^{10}$ & $3.6\times 10^{10}$ & $4.4\times 10^{2}$ & $2.1\times 10^{-3}$ & $1.5\times 10^{-6}$ \\
$$ & $\tau\bar{\tau}$ & $3.3$ & $1.8$ & $6.0\times 10^{-5}$ & $5.1\times 10^{10}$ & $2.9\times 10^{10}$ & $4.1\times 10^{2}$ & $5.1\times 10^{-4}$ & $3.7\times 10^{-7}$ \\
&&&&&&&&& \\[-0.1in]
\hline
&&&&&&&&& \\[-0.1in]
$2000$ & $b\bar{b}$ & $3.8$ & $3.3$ & $4.8\times 10^{-6}$ & $6.8\times 10^{11}$ & $6.9\times 10^{11}$ & $9.9\times 10^{3}$ & $5.4\times 10^{-2}$ & $3.8\times 10^{-5}$ \\
$$ & $W^{+}W^{-}$ & $3.3$ & $1.8$ & $4.8\times 10^{-5}$ & $6.5\times 10^{10}$ & $3.7\times 10^{10}$ & $4.7\times 10^{2}$ & $3.9\times 10^{-3}$ & $2.8\times 10^{-6}$ \\
$$ & $\tau\bar{\tau}$ & $3.3$ & $1.8$ & $6.0\times 10^{-5}$ & $5.1\times 10^{10}$ & $2.9\times 10^{10}$ & $4.4\times 10^{2}$ & $9.1\times 10^{-4}$ & $6.4\times 10^{-7}$ \\
&&&&&&&&& \\[-0.1in]
\hline
&&&&&&&&& \\[-0.1in]
$3000$ & $b\bar{b}$ & $3.8$ & $3.3$ & $5.4\times 10^{-6}$ & $6.1\times 10^{11}$ & $6.2\times 10^{11}$ & $1.0\times 10^{3}$ & $1.1\times 10^{-1}$ & $7.3\times 10^{-5}$ \\
$$ & $W^{+}W^{-}$ & $3.3$ & $1.8$ & $4.3\times 10^{-5}$ & $7.2\times 10^{10}$ & $4.0\times 10^{10}$ & $5.3\times 10^{2}$ & $9.9\times 10^{-3}$ & $6.8\times 10^{-6}$ \\
$$ & $\tau\bar{\tau}$ & $3.3$ & $1.8$ & $5.5\times 10^{-5}$ & $5.6\times 10^{10}$ & $3.2\times 10^{10}$ & $5.0\times 10^{2}$ & $2.2\times 10^{-3}$ & $1.5\times 10^{-6}$ \\
&&&&&&&&& \\[-0.1in]
\hline
&&&&&&&&& \\[-0.1in]
$5000$ & $b\bar{b}$ & $3.8$ & $3.3$ & $6.1\times 10^{-6}$ & $5.4\times 10^{11}$ & $5.5\times 10^{11}$ & $1.0\times 10^{3}$ & $2.6\times 10^{-1}$ & $1.7\times 10^{-4}$ \\
$$ & $W^{+}W^{-}$ & $3.6$ & $1.7$ & $3.9\times 10^{-5}$ & $8.2\times 10^{10}$ & $4.3\times 10^{10}$ & $5.9\times 10^{2}$ & $3.0\times 10^{-2}$ & $2.0\times 10^{-5}$ \\
$$ & $\tau\bar{\tau}$ & $3.6$ & $1.7$ & $4.7\times 10^{-5}$ & $6.8\times 10^{10}$ & $3.5\times 10^{10}$ & $5.7\times 10^{2}$ & $6.7\times 10^{-3}$ & $4.5\times 10^{-6}$ \\
&&&&&&&&& \\[-0.1in]
\hline
&&&&&&&&& \\[-0.1in]
$10000$ & $b\bar{b}$ & $3.8$ & $3.3$ & $6.0\times 10^{-6}$ & $5.5\times 10^{11}$ & $5.6\times 10^{11}$ & $1.2\times 10^{3}$ & $1.0\times 10^{0}$ & $6.7\times 10^{-4}$ \\
$$ & $W^{+}W^{-}$ & $3.3$ & $1.8$ & $2.9\times 10^{-5}$ & $1.1\times 10^{11}$ & $6.0\times 10^{10}$ & $8.3\times 10^{2}$ & $1.7\times 10^{-1}$ & $1.1\times 10^{-4}$ \\
$$ & $\tau\bar{\tau}$ & $3.3$ & $1.8$ & $3.2\times 10^{-5}$ & $9.6\times 10^{10}$ & $5.4\times 10^{10}$ & $8.9\times 10^{2}$ & $4.1\times 10^{-2}$ & $2.7\times 10^{-5}$ \\
\hline
\hline
\end{tabular}
\caption{Results after optimisation and unblinding for the angular separation $\Psi_{cut}$, the $90\%$ CL upper limit on the expected signal $\rm \mu^{90\%}$, the total averaged 
effective area $\rm \bar{A}_{eff}(M_{\rm WIMP}) = \sum_{i} \bar{A}_{eff}^{i}(M_{WIMP}) \times T_{eff}^{i}$ (with $i$ corresponding to a given period of the detector), the 
$90\%$ CL sensitivities on the neutrino+anti-neutrino flux at the Earth $\rm \overline{\Phi}_{\nu_{\mu}+\bar{\nu}_\mu}$, and the $90\%$ CL limits on the neutrino+anti-neutrino 
flux at the Earth $\rm \Phi_{\nu_{\mu}+\bar{\nu}_\mu}$, on the muon flux at the detector $\rm \Phi_{\mu}$ ($E_{\mu} > 1$ GeV), and on the spin-dependent and spin-independent 
WIMP-proton cross-sections $\rm \sigma_{p,SD}$ and $\rm \sigma_{p,SI}$ respectively. Results for $\rm M_{\rm WIMP}<500$ GeV are available in Table~\ref{tab:qresultsone}.
\label{tab:qresultstwo}}
\end{footnotesize}
\end{center}
\end{table*}

\end{document}